\documentclass[aps,pre,reprint,superscriptaddress,10pt]{revtex4-2}

\usepackage{graphicx,amsmath,amssymb,dcolumn,physics,color}

\begin{document}

\title{Tractable Model for Tunable Non-Markovian Dynamics}

\author{Matthew P.~Leighton}
\email{matthew.leighton@yale.edu}
\affiliation{Department of Physics and Quantitative Biology
Institute, Yale University, New Haven, CT 06511}

\author{Christopher W.~Lynn}
\email{christopher.lynn@yale.edu}
\affiliation{Department of Physics and Quantitative Biology
Institute, Yale University, New Haven, CT 06511}
\affiliation{Wu Tsai Institute, Yale University, New Haven, CT 06510}
\begin{abstract}
Non-Markovian dynamics are ubiquitous across physics, biology, and engineering. Yet our understanding of non-Markovian processes significantly lags that of simpler Markovian processes, due largely to a lack of tractable models. In this article, we present a minimal model of non-Markovian dynamics in which the current state copies past states with arbitrary history dependence. We show that many properties of this process can be studied analytically, providing insight into the relationships between history dependence, autocorrelations, and information-theoretic metrics like entropy and dynamical information. Strikingly, we find that autocorrelations can fail, even qualitatively, to capture the underlying dependencies. Ultimately, this model serves as a tractable sandbox for exploring non-Markovian dynamics.
\end{abstract}

\maketitle

\section{Introduction}

Non-Markovian stochastic processes are a ubiquitous feature of the natural world, spanning fields including biology~\cite{vroylandt2022likelihood,d2014mechanisms,zeraati2023intrinsic, cavanagh2020diversity,elgamel2023multigenerational,alba2020exploring, bialek2024long, barabasi2005origin}, language~\cite{shannon1948mathematical, wieczynski2025long}, and physics~\cite{breuer2016colloquium,de2017dynamics}. Such processes are defined by transition probabilities $p(x_t|x_{t-1},x_{t-2},\hdots )$, which encode the dependence of the current state $x_t$ on the history of the system. If $x_t$ depends only on the previous state $x_{t-1}$, then $p(x_t|x_{t-1},x_{t-2},\hdots ) = p(x_t | x_{t-1})$, and the dynamics are Markovian. Any additional dependencies produce non-Markovian dynamics. For example, if each state depends on the previous $N$ states, then $p(x_t|x_{t-1},\hdots ) = p(x_t|x_{t-1},\hdots ,x_{t-N})$, and the dynamics are of order $N$. More generally, these non-Markovian dependencies can stretch infinitely far into the past.

The study of non-Markovian processes presents significant technical and practical challenges. Finite-order non-Markovian dynamics can be mapped onto Markovian dynamics; however, at the cost of an exponential explosion of states. Even this strategy fails for infinite-order non-Markovian dynamics, where dependencies reach arbitrarily far into the past. As a result, tractable models of non-Markovian processes are rare~\cite{boguna2013simulating,da2015two}, thus limiting our understanding of dynamics with long-range dependencies.

Perhaps the most prominent examples of non-Markovian processes are hidden Markov models, where one observes only a subset of the degrees of freedom in an underlying Markov process~\cite{rabiner2002tutorial, zucchini2009hidden}. More generally, any coarse-graining that destroys information about the state of a Markovian process can lead to dynamics that appear non-Markovian~\cite{agon2018coarse, strasberg2019non, schwarz2024mind}. Beyond finite-order and hidden Markov models lie non-Markovian systems with infinite-order history dependence~\cite{mochihashi2007infinite,marzen2016predictive}. Among the most well-studied examples is the Elephant Random Walk (ERW)~\cite{schutz2004elephants}, which, as detailed below, arises as a special case of our proposed model. However, there exist few other examples of analytically tractable processes with arbitrarily long history dependence~\cite{da2015two,georgiou2015solvable,saha2022random, boyer2025intermittent}. Since non-Markovian dynamics can defy Markovian intuitions, developing a minimal framework for exploring long-range dependencies is of critical importance.

In this article, we introduce a model of non-Markovian dynamics that is simple enough to allow analytic solutions yet flexible enough to study arbitrary memory into the past. We compute autocorrelations and information-theoretic quantities for several key classes of history dependence. In doing so, we gain new insights into the relationships between history dependence, common statistics such as autocorrelations, and more advanced measures of entropy and dynamical information.

\section{Non-Markovian Copy Model}

\begin{figure}[h]
    \centering
    \includegraphics[width =\linewidth]{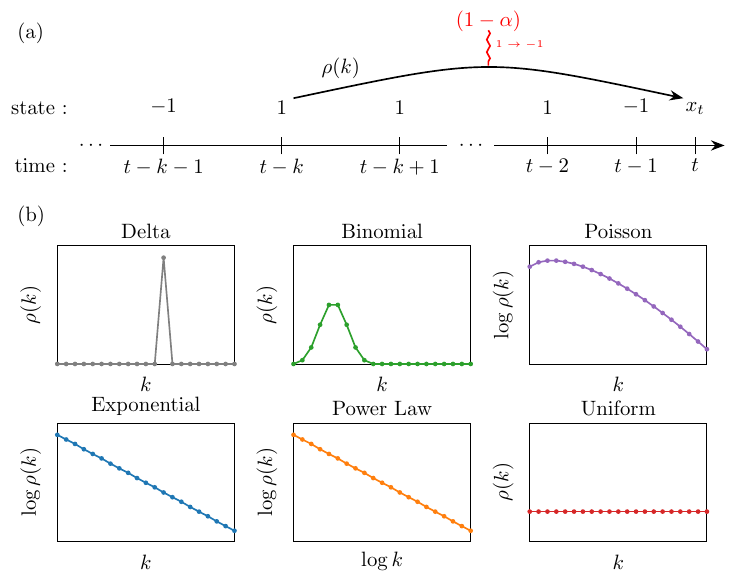}
    \caption{Schematic illustrating the non-Markovian copy process. (a) At each time $t$, the state $k$ steps in the past $x_{t-k}$ is selected with probability $\rho(k)$ and then copied to the current state $x_t$ with probability $\alpha$ (yielding error probability $1-\alpha$). (b) The strength of the $k^\text{th}$-order dependence is defined by the distribution $\rho(k)$, which can be tuned arbitrarily to produce different non-Markovian dynamics of both finite and infinite order.}
    \label{fig:fig1}
\end{figure}

Consider a discrete-time stochastic process [illustrated in Fig.~\ref{fig:fig1}(a)], where at each time $t$ the state takes a binary value $x_t= \pm1$. Given the history of the system, $x_t$ is generated by first selecting the state $k$ steps in the past with probability $\rho(k)$ and then copying this state with probability $\alpha$ (yielding $x_t = x_{t-k}$) or failing with probability $1-\alpha$ (yielding $x_t = -x_{t-k}$). In this way, the distribution $\rho(k)$, which can be tuned arbitrarily, defines the strength of non-Markovian dependencies. Given the full history, $x_t$ is conditionally-distributed according to 
\begin{equation}\label{eq:condprob}
p(x_t|x_{t-1},\hdots ) = \frac{1}{2} + \frac{1}{2}\sum_{k=1}^\infty (2\alpha-1) \rho(k)x_{t-k}x_t.
\end{equation}
In what follows, we focus primarily on the behavior of the system when $\alpha>0.5$, such that successful copying is more likely than failure. We also assume that the system is initialized randomly such that $\langle x_t\rangle = 0$ for all $t$. 

We consider several qualitatively different history dependence distributions $\rho(k)$ [Fig.~\ref{fig:fig1}(b)]. The simplest dynamics are defined by delta-distributed history dependence
\begin{equation}
\label{eq:rho_delta}
\rho(k)=\delta(k-N),
\end{equation}
where each state copies the $N^\text{th}$ state in the past. Another finite-order process is given by the binomial distribution
\begin{equation}
\label{eq:rho_binom}
\rho(k) = {N-1 \choose k-1}q^{k-1}(1-q)^{N-k},
\end{equation}
which is defined on $1\le k \le N$, with a shape parameter $0\leq q\leq1$ that shifts the peak of the distribution either closer to the present ($q\to 0$) or further into the past ($q\to 1$).

In addition to the delta and binomial processes (which are of finite order $N$), we also investigate several distributions that give rise to infinite-order non-Markovian dynamics. As an infinite-order counterpart to the binomial distribution, we use a shifted Poisson distribution
\begin{equation}
\rho(k) = \frac{\lambda^{k-1}e^{-\lambda}}{(k-1)!},
\end{equation}
where $\lambda>0$ controls the position of the distribution's peak. The Poisson distribution decays superexponentially with $k$. For exponentially-decaying history dependence, we study the geometric distribution
\begin{equation}
\rho(k) = q (1-q)^{k-1},
\end{equation}
where $0< q \leq 1$ a shape parameter which controls how quickly the memory decays. To investigate power-law dependencies, we consider the zeta distribution
\begin{equation}
\rho(k) = \frac{k^{-s}}{\zeta(s)},
\end{equation}
where $\zeta(s)$ is the Riemann zeta function, and we require $s>1$ so the distribution is properly normalized.

Finally, in the limit where all dependencies of all orders are of equal strength, the distribution $\rho(k)=1/t$ is uniform. This process is equivalent to the well-studied elephant random walk (ERW), which defines a one-dimensional random walk where left/right steps are taken at a rate proportional to the previous number of steps in that direction. In this way, the process features memory arbitrarily far into the past (since ``elephants never forget"). Analytic solutions for many properties of interest have been derived, including autocorrelations  and limiting distributions of various observables~\cite{kenkre2007analytic, baur2016elephant, coletti2017central, laulin2022introducing, gut2023elephant}. Unlike the other distributions described above, the uniform distribution must be time-dependent to ensure normalization. As a result, the conditional distribution $p(x_t|x_{t-1},\hdots )$ does not reach a time-independent steady state.

\section{Correlations, Entropy, and Information}
\label{sec:generalresults}

In this section, we derive general results for the non-Markovian copy process that are independent of the specific history dependence $\rho(k)$. Our ultimate goal is to characterize autocorrelations and information-theoretic quantities like entropy and information. Toward this end, we also introduce additional quantities that aid in our exploration.

\subsection{Autocorrelations}
Autocorrelations are commonly used to study non-Markovian dynamics~\cite{linkenkaer2001long,barabasi2005origin,zeraati2023intrinsic,alba2020exploring, bialek2024long}, often relying on the heuristic that the long-time scaling behavior of the autocorrelation function can be used to learn about the underlying history dependence. While Markovian dynamics generally feature exponentially-decaying autocorrelations, and non-Markovian dynamics often feature power-law autocorrelations, this heuristic does not always hold~\cite{leighton2025main}. Thus an important question for the study of this non-Markovian copy process is how the autocorrelations relate to the underlying history dependence $\rho(k)$.

The autocorrelation function $C(k)$ is defined as
\begin{equation}
C(k) \equiv \left\langle x_t x_{t-k}\right\rangle - \left\langle x_t\right\rangle\left\langle x_{t-k}\right\rangle.
\end{equation}
Note that for all choices of $\rho(k)$, the process is unbiased such that $\left\langle x_t\right\rangle=0$, and thus the second term vanishes. Here and throughout, angle brackets with no subscript denote averages over the steady-state distribution $p(x_t,x_{t-1},\hdots )$. When a subscript is present, it denotes the probability distribution over which the average is to be taken. The correlations can be written as a composite average:
\begin{equation}\label{eq:Ckexpansion}
C(k) = \left\langle \left\langle x_t x_{t-k}\right\rangle_{p(x_t,x_{t-k}|k')}\right\rangle_{\rho(k')},
\end{equation}
where $k'$ is a dummy variable representing the time-delay selected for copying at time $t$, so that $p(x_t,x_{t-k}|k')$ is the joint probability of $x_t$ and $x_{t-k}$ given that the lag $k'$ is chosen at time $t$. This formulation allows us to relate the correlation between $x_t$ and $x_{t-k}$ to that between $x_{t-k'}$ and $x_{t-k}$. We do this by evaluating the interior average, the correlation of $x_t$ and $x_{t-k}$ given that the lag $k'$ is chosen at time $t$, as
\begin{subequations}
\begin{align}
\left\langle x_t x_{t-k}\right\rangle_{p(x_t,x_{t-k}|k')}  = & \alpha \langle x_{t-k'}x_{t-k}\rangle \nonumber\\
& - (1-\alpha)\langle x_{t-k'}x_{t-k}\rangle\\
=& (2\alpha-1)C(k'-k).
\end{align}
\end{subequations}
Here we have used the fact that $C(k'-k)=C(k-k')$ for this process, since we did not require $k'\leq k$. Inserting into Eq.~\eqref{eq:Ckexpansion}, we find that the autocorrelations obey the recurrence relation
\begin{equation}
\label{autocorrelationrecurrence}
C(k) = (2\alpha-1)\sum_{k'=1}^\infty C(k-k')\rho(k').
\end{equation}
As detailed below, for some dependencies $\rho(k)$, this recurrence relation [combined with $C(0)=1$] is sufficient to derive an analytic solution for $C(k)$.

\subsection{Generating functions}

When full solutions are not available, to determine the long-time scaling of the autocorrelations, it will prove useful to examine the generating functions
\begin{align}
\tilde{\rho}(z) &\equiv \sum_{k=1}^\infty \rho(k) z^k,\\
\tilde{C}(z) &\equiv \sum_{k=0}^\infty C(k) z^k,
\end{align}
of the dependence distribution $\rho(k)$ and autocorrelations $C(k)$, respectively. Multiplying both sides of the recurrence relation [Eq.~\eqref{autocorrelationrecurrence}] by $z^k$ and summing over $k$, we obtain a direct relationship between $\tilde{\rho}$ and $\tilde{C}$:
\begin{subequations}
\label{eq:eq12}
\begin{align}
\tilde{C}(z)-1 & = \sum_{k=1}^\infty z^k C(k)\\
& = (2\alpha-1)\sum_{k=1}^\infty\sum_{n=1}^\infty \rho(n)C(k-n)z^k\\
& = (2\alpha-1)\sum_{n=1}^\infty \rho(n) z^n\sum_{k'=1-n}^\infty C(k')z^{k'}\\
& = (2\alpha-1)\tilde{\rho}(z)\tilde{C}(z) + (2\alpha-1)A(z),
\end{align}
\end{subequations}
where we define the polynomial
\begin{equation}
A(z) \equiv \sum_{n=1}^\infty\sum_{j=1}^{n-1}\rho(n)C(j)z^{n-j},
\end{equation}
which we note is analytic for all $z$. Equation~\eqref{eq:eq12} can then be solved to obtain
\begin{equation}
\label{eq:Cz}
\tilde{C}(z) = \frac{1+(2\alpha-1)A(z)}{1 - (2\alpha-1)\tilde{\rho}(z)}.
\end{equation}

In terms of this generating function, the autocorrelations are given by Cauchy's integral formula:
\begin{equation}\label{eq:cauchy}
C(k) = \frac{1}{2\pi i} \oint_\gamma \frac{\tilde{C}(z)}{z^{k+1}}\mathrm{d}z.
\end{equation}
As $k\to\infty$, this integral is dominated by the pole or other discontinuity of $\tilde{C}(z)$ with $|z|$ closest to $1$. This allows the long-time scaling of the autocorrelations to be characterized using the residue theorem~\cite{conway2012functions}.

\subsection{Collective variables}

In the investigation of the copy process, an important quantity is the running average of past states,
\begin{equation}
Y_k(t) \equiv \sum_{j=1}^k \rho(j) x_{t-j},
\end{equation}
which we refer to as the collective variable. In the limit of infinite history, we also define
\begin{equation}\label{eq:Ytdef}
Y(t) \equiv \lim_{k\to\infty}Y_k(t).
\end{equation}
Note that while $Y_k(t)$ depends on time for a given trajectory, the distribution of $Y_k(t)$ converges to a time-independent steady state. When considering distribution-level quantities we will thus generally omit the time-dependence.

Using the collective variable, the conditional distribution of the current state $x_t$ given a finite history $\{x_{t-1},\hdots ,x_{t-k}\}$ simplifies dramatically,
\begin{subequations}
\begin{align}
p(x_t|x_{t-1},\hdots ,x_{t-k}) & = \left\langle p(x_t|x_{t-1},\hdots )\right\rangle_{p(x_{t-k-1},\hdots )}\\
& = \frac{1}{2} + \frac{2\alpha-1}{2} Y_k(t)x_t.
\end{align}
\end{subequations}
Above, we marginalize over the prior history $\{x_{t-k-1},\hdots \}$ and use the fact that $\langle Y_k\rangle=0$ for all $k$. In fact, since $\langle x_t\rangle=0$, it follows that all odd moments of $Y_k$ are zero. Thus, all information from the first three moments of the distribution of $Y_k$ is contained in the variance,
\begin{subequations}
\label{eq:Ykexpansion}
\begin{align}
\left\langle Y_k^2\right\rangle & = \left\langle \left( \sum_{j=1}^k \rho(j)x_{t-j}\right)^2\right\rangle\\
& = \sum_{j=1}^k \rho(j)^2\left\langle x_{t-j}^2\right\rangle + 2\sum_{1\leq j<l\leq k} \rho(j)p(l) \left\langle x_{t-j}x_{t-l}\right\rangle\\
& = \sum_{j=1}^k \rho(j)^2 + 2\sum_{1\leq j<l\leq k} \rho(j)p(l) C(l-j).
\end{align}
\end{subequations}
This provides a simple method for computing the variance of $Y_k$ directly from the history dependence $\rho(k)$ and autocorrelations $C(k)$.

\subsection{Entropy}

Information theory is the natural language with which to quantify history dependence in stochastic processes. Our uncertainty about the state $x_t$ given the history $\{x_{t-1},\hdots\}$ is quantified by the entropy rate~\cite{shannon1948mathematical}
\begin{equation}
h_\infty \equiv H[x_t|x_{t-1},\hdots ] = \langle -\ln p(x_t|x_{t-1},\hdots)\rangle.
\end{equation}
Given only a finite history up to $k$ steps into the past, $\{x_{t-1},\hdots,x_{t-k}\}$, our uncertainty about the state $x_t$ is instead given by the conditional entropy
\begin{equation}
h_k \equiv H[x_t|x_{t-1},\hdots ,x_{t-k}],
\end{equation}
such that $h_\infty = \lim_{k\to\infty}h_k$. When the past is entirely unknown, our uncertainty is simply the marginal entropy
\begin{equation}
h_0 \equiv H[x_t].
\end{equation}
For the copy process, because $\langle x_t \rangle = 0$, the marginal entropy is $h_0 = 1$~bit for all choices of $\rho(k)$ and $\alpha$. Since additional knowledge of the past can only reduce our uncertainty, the conditional entropy forms a hierarchy for increasing $k$,
\begin{equation}
\label{eq:hier}
h_0\geq h_1\geq\hdots\geq h_k\geq\hdots h_\infty \ge 0.
\end{equation}

We can bound and approximate these conditional entropies using the collective variable $Y_k$. The $k^\mathrm{th}$-order conditional entropy, for example, can be expanded as
\begin{subequations}
\label{eq:hk_approx}
\begin{align}
& h_k = \\
& -\left\langle \left(\frac{1}{2} + \frac{1}{2}(2\alpha-1)Y_k\right)\ln\left(\frac{1}{2} + \frac{1}{2}(2\alpha-1)Y_k\right) \right.\nonumber \\
& \left.+ \left(\frac{1}{2} - \frac{1}{2}(2\alpha-1)Y_k\right)\ln\left(\frac{1}{2} - \frac{1}{2}(2\alpha-1)Y_k\right)\right\rangle_{p(Y_k)}\nonumber\\
& = \ln2 -  \sum_{n=1}^\infty \frac{1}{2n(2n-1)}(2\alpha-1)^{2n} \left\langle Y_k^{2n}\right\rangle,
\end{align}
\end{subequations}
where in the last line we expand around $Y_k=0$. Note that since $|Y_k|\leq1$, higher-order terms in this series will become increasingly small. We can therefore approximate $h_k$ by truncating the infinite series at finite $n$. Compared to the brute force strategy of estimating the full conditional probabilities $p(x_t|x_{t-1},\hdots,x_{t-k})$, this presents a much more efficient method to compute conditional entropies from simulations.

We can also use Eq.~\eqref{eq:hk_approx} to derive bounds on the conditional entropy based on simple statistics of the collective variable $Y_k$. Noting that Jensen's inequality~\cite{Cover2006_Elements} implies $\langle Y_k^{2n}\rangle \geq \langle Y_k^2\rangle^n$ for all $n\geq1$, we can upper-bound the conditional entropy as
\begin{subequations}
\begin{align}
h_k & \leq \ln2 -  \sum_{n=1}^\infty \frac{1}{2n(2n-1)}(2\alpha-1)^{2n} \left\langle Y_k^{2}\right\rangle^n \label{eq:h_kseries} \\
& = \ln2 - \Phi\left((2\alpha-1)\sqrt{\langle Y^2_k\rangle}\right),
\end{align}
\end{subequations}
where
\begin{equation}
\Phi(x) = \frac{1}{2}\left[(1+x)\ln(1+x) + (1-x)\ln(1-x)\right].
\end{equation}
An accompanying lower bound can be derived by noting $\langle Y_k^{2n}\rangle\leq\langle Y_k^2\rangle$ for all $n\geq 1$, so that
\begin{equation}\label{eq:hkLB}
h_k \geq \ln2 - \Phi\left(2\alpha-1\right)\langle Y^2_k\rangle.
\end{equation}
Using these bounds, we can constrain all $h_k$ from simulation data using only the fidelity $\alpha$ and variance of the collective variable $\langle Y^2_k\rangle$. In particular, these gives simple lower and upper bounds on the entropy rate:
\begin{equation}
\label{eq:hinftybounds}
\begin{aligned}
\ln2 - \Phi\left(2\alpha-1\right)\langle Y^2\rangle & \leq h_\infty \\
& \leq \ln2 - \Phi\left((2\alpha-1)\sqrt{\langle Y^2\rangle}\right).
\end{aligned}
\end{equation}

\subsection{Dynamical information}

With tools to quantify entropy, we can now ask how much the past reduces our uncertainty about the future. Given knowledge of the entire past, our uncertainty about the next state is reduced from the marginal entropy $h_0$ to the entropy rate $h_\infty$. Thus, we have gained an amount of information
\begin{equation}
I_\mathrm{tot} \equiv h_0 - h_\infty \ge 0,
\end{equation}
which we refer to as the \emph{dynamical information}~\cite{leighton2025main}. 

At a more granular level, with knowledge of the previous $k-1$ states $\{x_{t-1},\hdots,x_{t-k+1} \}$, we can ask how much one more state $x_{t-k}$ reduces our uncertainty about the next state $x_t$. This is precisely the difference in conditional entropies,
\begin{equation}
I_k \equiv h_{k-1} - h_k \ge 0,
\end{equation}
which we refer to as the $k^\text{th}$-order dynamical information. If $h_{k-1} = h_k$, then $x_k$ is redundant in the sense that the dynamics are entirely determined by lower-order dependencies, and thus $I_k = 0$. by contrast, if $h_{k-1} > h_k$, then $x_k$ provides new information $I_k > 0$ about the future of the system. This reduction in uncertainty is equivalent mutual information between $x_t$ and $x_{t-k}$ conditioned on the intervening history,
\begin{equation}
I_k = I[x_t;x_{t-k}\, |\, x_{t-1},\hdots,x_{t-k+1}].
\end{equation}

The hierarchy in Eq.~\eqref{eq:hier} tells us that the total dynamical information $I_\mathrm{tot}$ decomposes into non-zero contributions from each order $k$,
\begin{align}
I_\mathrm{tot} = I_1 + I_2 + \cdots = \sum_{k = 1}^\infty I_k.
\end{align}
This decomposition reveals how dependencies of all orders combine to constrain non-Markovian dynamics~\cite{leighton2025main}. For example, if the dynamics are of finite order $N$, then $I_k = 0$ for all $k > N$, and the dynamical information decomposes into a sum of $N$ parts,
\begin{equation}
I_\text{tot} = I_1 + \cdots + I_N.
\end{equation}
It will also become useful to separate the total dynamical information $I_\text{tot} = I_1 + I_{>1}$ into the Markovian information $I_1$ and the combined information from non-Markovian dependencies,
\begin{equation}
I_{>1} \equiv \sum_{k = 2}^\infty I_k.
\end{equation}

Applying the bounds in Eq.~\eqref{eq:hinftybounds}, we derive upper and lower bounds on the total dynamical information:
\begin{subequations}
\begin{align}
I_\mathrm{tot}& \geq \Phi\left((2\alpha-1)\sqrt{\langle Y^2\rangle}\right),\label{eq:dynamicalinfoapprox}\\
I_\mathrm{tot} & \leq \Phi\left(2\alpha-1\right)\langle Y^2\rangle.\label{eq:ItotUB}
\end{align}
\end{subequations}
These bounds allow us to constrain $I_\mathrm{tot}$ from simulation data using only the fidelity $\alpha$ and variance of the collective variable $\langle Y^2\rangle$.

We then turn to the $k^\mathrm{th}$-order dynamical information. Using our results for the conditional entropy $h_k$, we can bound and approximate the dynamical information $I_k$. Taking the difference between subsequent conditional entropies in Eq.~\eqref{eq:hk_approx}, the $k^\mathrm{th}$-order dynamical information can be written
\begin{equation}
\label{eq:dynamicalinfoseries}
I_k = \sum_{n=1}^\infty \frac{(2\alpha-1)^{2n}}{2n(2n-1)} \left(\left\langle Y_k^{2n}\right\rangle - \left\langle Y_{k-1}^{2n}\right\rangle\right).
\end{equation}
This series provides an efficient method for computing the dynamical information from simulation data. In practice, we find that this series converges to within $<1\%$ using only the first $n=3$ terms for all distributions $\rho(k)$ considered.

We can approximate the large-$k$ scaling of the dynamical information using the first-order truncation of the series~\eqref{eq:dynamicalinfoseries}. Inserting the expansion for the variance of the collective variable in Eq.~\eqref{eq:Ykexpansion}, we obtain
\begin{subequations}
\begin{align}
I_k
& \approx \frac{1}{2}(2\alpha-1)^2 \left(\left\langle Y_k^{2}\right\rangle - \left\langle Y_{k-1}^{2}\right\rangle \right)\\
& = \frac{1}{2}(2\alpha-1)^2\left[\rho(k)^2 + 2C(k) \sum_{j=1}^{k-1}\rho(j) \rho(j+k)) \right].
\end{align}
\end{subequations}
For $\alpha\geq 0.5$, we have $C(k)\geq 0$ for all $k$, which leads to an approximate lower bound,
\begin{equation}
\label{eq:dyninfobound}
I_k\gtrsim \frac{1}{2}(2\alpha-1)^2 \rho(k)^2.
\end{equation}
Similarly, using the recurrence relation for $C(k)$ [Eq.~\eqref{autocorrelationrecurrence}] and requiring that $C(k)$ decays monotonically, an upper bound is 
\begin{equation}
I_k\lesssim \frac{1}{2}(2\alpha-1)^2 \rho(k)^2 + (2\alpha-1) \rho(k)C(k).
\end{equation}
When $C(k)$ decays faster than $\rho(k)$ (as in many, but not all, of the cases considered in this article), we then expect the limiting behavior as $k\to\infty$ to be
\begin{equation}
I_k \sim \frac{1}{2}(2\alpha-1)^2\rho(k)^2.
\end{equation}
This approximate scaling establishes that, for many choices of $\rho(k)$, the dynamical information $I_k$ correctly captures the underlying history dependence $\rho(k)$. More generally, the dynamical information may decay slower than $\rho(k)^2$, but never faster. As we will see in Sec.~\ref{sec:infiniteorder}, this is not always the case for autocorrelations.

\begin{figure*}[t!]
    \centering
    \includegraphics[width =0.95\linewidth]{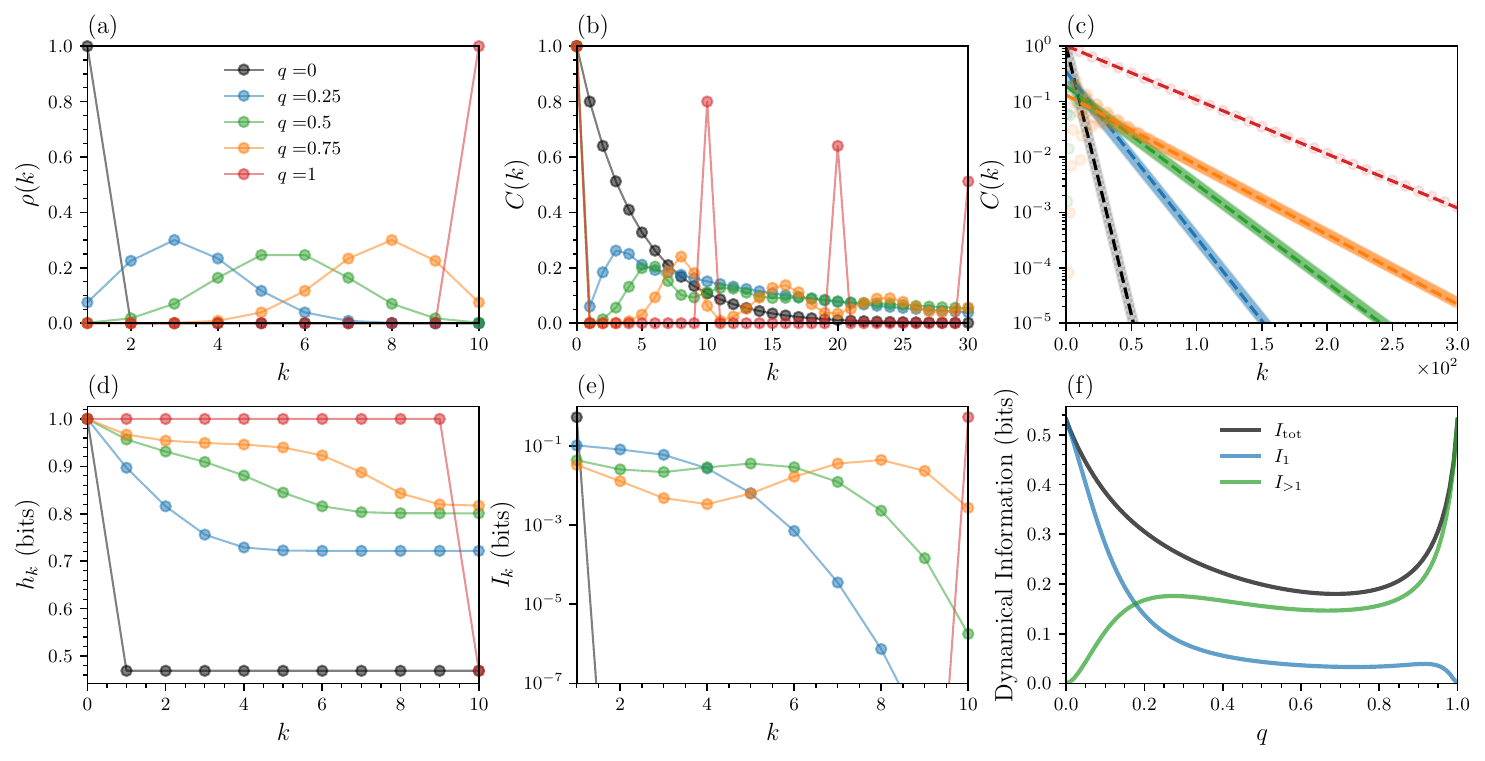}
    \caption{Copy process with finite-order history dependence. (a) Binomial history dependence distributions $\rho(k)$ [Eq.~\eqref{eq:rho_binom}] with order $N=10$ and shape parameters $q = \{0,0.1,0.5,0.9,1.0\}$. Parameters $q = 0,1$ are equivalent to delta history dependence [Eq.~\eqref{eq:rho_delta}] with $N = 1,10$, respectively. (b-c) Autocorrelations $C(k)$ with points reflecting numerical values and dashed lines in (c) depicting theoretical predictions for exponential scaling Eq.~\eqref{eq:binomialCk}. (d-e) Conditional entropy $h_k$ (d) and dynamical information $I_k$ (e) as functions of order $k$. (f) Total, Markovian, and non-Markovian dynamical information versus shape parameter $q$. Values in (d-f) are computed exactly, and $\alpha = 0.9$ for all analyses.}
    \label{fig:fig2}
\end{figure*}

\section{Finite-Order Processes}
\label{sec:finiteorder}

We are now prepared to investigate the properties of the copy model with specific history dependencies $\rho(k)$. For simplicity, we begin with dynamics of finite order $N$, such that $\rho(k)=0$ for $k>N$, before studying infinite-order dynamics in the following section. Even for these relatively simple finite-order processes, our results illustrate the possibility for counterintuitive behavior of both autocorrelations and information-theoretic measures of history dependence.

\subsection{Delta history dependence}

We first consider a memory that contains only the $N^\text{th}$ state in the past, such that
\begin{equation}
\label{eq:delta}
\rho(k) = \delta(k-N).
\end{equation}
If $N = 1$, then the dynamics are Markovian. More generally, Eq.~\eqref{eq:delta} defines an $N^\text{th}$-order process, but with no dependence on the intervening history. In Fig.~\ref{fig:fig2}(a), we illustrate two examples corresponding to $N=1$ and $N=10$. The autocorrelation can be evaluated directly for this process using Eq.~\eqref{autocorrelationrecurrence}, yielding
\begin{equation}
C(k) = \begin{cases} (1-\alpha)^k & k/N\in\mathbb{N}\\
0 & k/N\notin \mathbb{N}, \end{cases}
\end{equation}
where $\mathbb{N}$ is the set of natural numbers. The correlations therefore decay exponentially over time, with non-zero values only at multiples of $N$ [Figs.~\ref{fig:fig2}(b-c)].

The conditional entropy can similarly be computed exactly. Since the state $x_t$ depends only on the state $x_{t-N}$, the entropy rate is simply the binary entropy for a Bernoulli process with probability $\alpha$, $h_\infty = -\alpha\ln\alpha - (1-\alpha)\ln(1-\alpha)$. More generally, the conditional entropy $h_k$ is given by
\begin{equation}
h_k = \begin{cases}\ln2, & k<N\\
-\alpha\ln\alpha - (1-\alpha)\ln(1-\alpha), & k\geq N.
\end{cases}
\end{equation}
Thus, the only drop in conditional entropy arises due to the $N^\mathrm{th}$ state in the past.

The dynamical information can then be computed from the conditional entropies, which reveals that only the $N^\text{th}$-order dynamical information is non-zero:
\begin{equation}
I_k = \begin{cases}
\ln2 + \alpha\ln\alpha + (1-\alpha)\ln(1-\alpha), & k=N\\
0, & \mathrm{otherwise}.
\end{cases}
\end{equation}
Thus, the dynamical information accurately captures the delta-function history dependence, picking out exactly the timepoint that is being copied while the rest of the history provides no information [Fig.~\ref{fig:fig2}(e)].

\subsection{Binomial history dependence}

For a more complicated example of $N^\text{th}$-order history dependence, we consider the binomial distribution 
\begin{equation}
\rho(k) = {N-1 \choose k-1}q^{k-1}(1-q)^{N-k},
\end{equation}
where $q\in[0,1]$ is a shape parameter and we focus on the case $N = 10$. As shown in Fig.~\ref{fig:fig2}(a), for small $q$ the distribution concentrates around $k=1$, while for large $q$ the distribution concentrates near $k=N$. Thus, the limits $q=0$ and $q=1$ are equivalent to the delta-distributed dependence discussed above with $N = 1$ and $N = 10$, respectively.

To study the autocorrelations, we simulate the copy process and compute $C(k)$ numerically. Due to the non-monotonic history dependence $\rho(k)$, the autocorrelations oscillate non-monotonically for small $k$ [Fig.~\ref{fig:fig2}(b)]. For large $k$, however, the correlations decay exponentially [Fig.~\ref{fig:fig2}(c)]. To extract the asymptotic (large--$k$) scaling analytically, we apply the residue theorem to the pole of $\tilde{C}(z)$ nearest to $z=1$. The generating function for the history dependence distribution is $\tilde{\rho}(z) = z(1-q+qz)^N$, so expanding the denominator of Eq.~\eqref{eq:Cz} around $z=1$ allows us to identify the pole closest to $z=1$,
\begin{equation}
z_c = 1 + \frac{2(1-\alpha)}{(2\alpha-1)(1+qN)}.
\end{equation}
Thus, near $z=1$ we can rewrite the autocorrelation generating function as
\begin{equation}
\tilde{C}(z)\propto\frac{1}{1-z/z_c}.
\end{equation}
The residue theorem then provides the asymptotic form for the autocorrelations,
\begin{subequations}
\label{eq:binomialCk}
\begin{align}
C(k\gg1) & = \lim_{z\to z_c}(z_c-z)\frac{\tilde{C}(z)}{z^{k+1}}\\
& = \frac{1 + (2\alpha-1)A(z_c)}{z_c^k}\\
& \propto \left(1 + \frac{2(1-\alpha)}{(2\alpha-1)(1+qN)}\right)^{-k}.
\end{align}
\end{subequations}
This analytic prediction matches simulations [Fig.~\ref{fig:fig2}(c)], confirming that the autocorrelations decay exponentially in the large--$k$ limit.

We next examine information-theoretic measures of history dependence. The conditional entropy $h_k$ decreases monotonically as more history is taken into account, reaching the entropy rate $h_\infty$ at $k=10$ [Fig.~\ref{fig:fig2}(d)], which is required since the dynamics are $N^\mathrm{th}$-order. As a function of the shape parameter $q$, the entropy rate $h_\infty$ is non-monotonic, exhibiting minimum values in the limits $q=0,1$ [where $\rho(k)$ is highly concentrated] and a maximum at intermediate $q$ [where $\rho(k)$ is more diffuse]. The history dependence is captured by the $k^\mathrm{th}$-order dynamical information $I_k$ [Fig.~\ref{fig:fig2}(e)]. In the two delta-distributed cases, the dynamical information is only non-zero for a single order ($k = 1$ for $q = 0$ and $k = 10$ for $q = 1$), reflecting the fact that all of the historical dependence is concentrated on a single state in the past.  Meanwhile, for intermediate $q$, we find that $I_k$ is distributed throughout the history in a manner which roughly matches the true dependencies $\rho(k)$. For $k>N$, when $\rho(k) = 0$, the dynamical information also vanishes ($I_k = 0$). 

Finally, we seek to quantify the degree to which the dynamics of the copy process are non-Markovian. To do so, we explore the total dynamical information ($I_\text{tot}$) and its decomposition into Markovian ($I_1$) and non-Markovian ($I_{>1}$) [Fig.~\ref{fig:fig2}(f)]. Recall that the Markovian and non-Markovian information sum to the total dynamical information, $I_\mathrm{tot} = I_1 + I_{>1}$. For $q=0$, the process is fully Markovian, such that $I_\mathrm{tot} = I_1$. For $q=1$, the dynamics have no dependence on the previous state, so that $I_1=0$ and $I_\mathrm{tot} = I_{>1}$. Between these extremes, intermediate $q$ produces a mixture of Markovian and non-Markovian information, with the proportion $I_{>1}/I_\text{tot}$ increasing with $q$. This leads to a local maximum of non-Markovian information at $q\approx 0.25$. These results demonstrate that the total dynamical information $I_\text{tot}$, and even the non-Markovian information $I_{>1}$, need not increase monotonically as the history dependence $\rho(k)$ reaches further into the past.

\section{Infinite-Order Processes}
\label{sec:infiniteorder}

We now investigate non-Markovian processes of infinite order, which have dependencies $\rho(k)>0$ for all orders $k$. Our analysis is focused in particular on understanding the relationships between the asymptotic scaling of history dependence, autocorrelations, and dynamical information. We organize our investigation to progress from short- to long-range dependencies, advancing from superexponential (Poisson) to exponential (geometric) to power-law (zeta) to uniform (ERW) scaling.

For all processes other than the ERW, we compute both the autocorrelations and the dynamical information from simulation data. Specifically, we use simulations to estimate the moments of $Y_k$ and compute $I_k$ using the infinite series Eq.~\eqref{eq:dynamicalinfoseries} truncated to third order. We find that this series converges to within 1\% by the third term for all parameter regimes considered in this article.

\begin{figure*}[t!]
    \centering
    \includegraphics[width =0.95\linewidth]{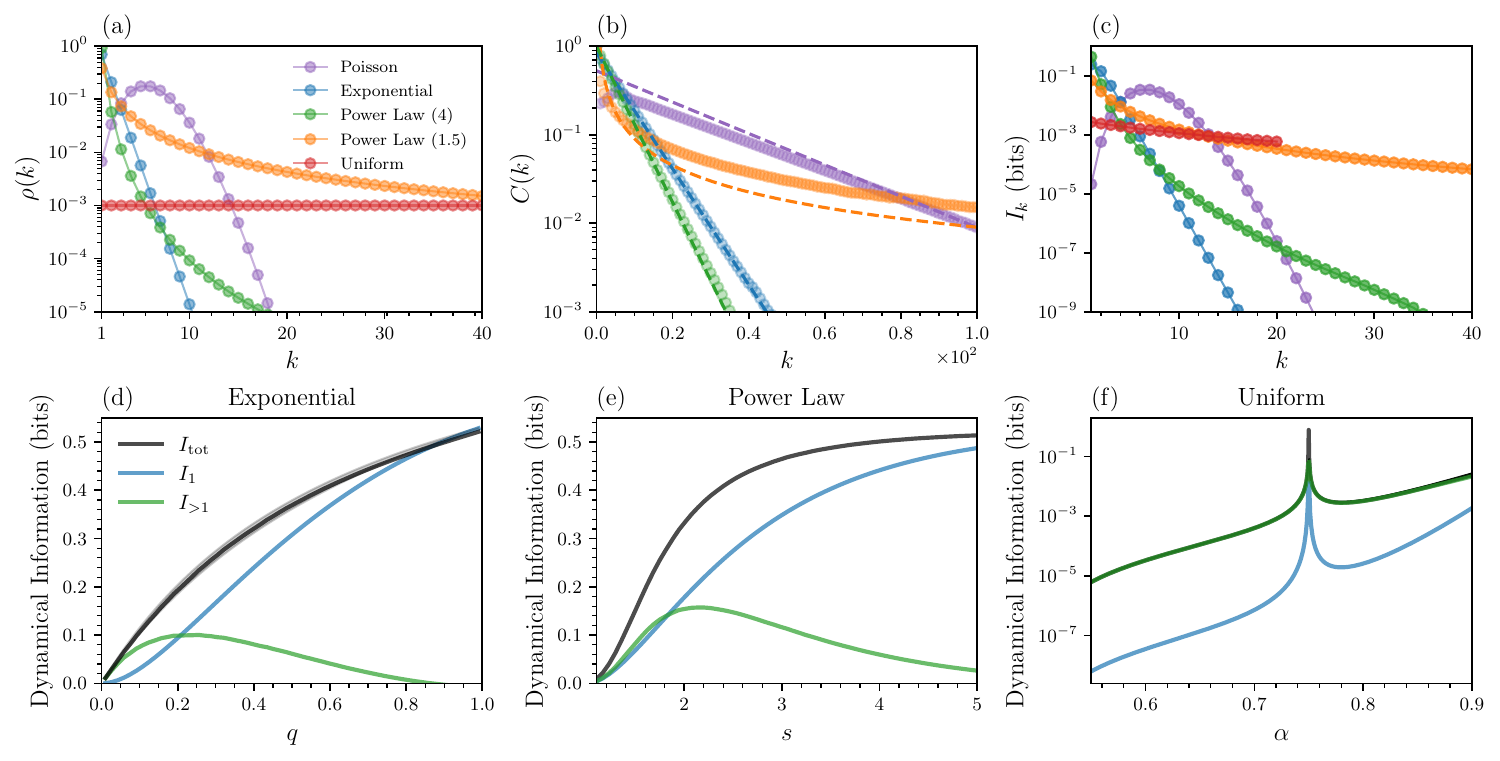}
    \caption{Copy process with infinite-order history dependence. (a) History dependencies $\rho(k)$ defined by Poisson ($\lambda=5$, purple), exponential ($q=0.7$, blue), power-law ($s=4$, green; $s=1.5$, orange), and uniform ($t=1000$, red) distributions. (b) Autocorrelations computed from simulations (points), with dashed lines representing analytic calculations from Eq.~\eqref{eq:poissonCk} (Poisson), Eq.~\eqref{eq:expCk} (exponential), and Eq.~\eqref{eq:zetaCk} (power-law). For Poisson, we use exponential scaling from Eq.~\eqref{eq:poissonCk} and match the intercept to simulations at $k=100$. (c) Dynamical information $I_k$ computed from simulations using the first three terms of Eq.~\eqref{eq:dynamicalinfoseries}. For ERW, $I_k$ is computed using the recursive algorithm in Appendix C up to $k=20$ (due to computational limitations). (d-f) Total, Markovian, and non-Markovian dynamical information for exponential dependence with varying $q$ (d), power-law dependence with varying $s$ (e), and uniform dependence with varying $\alpha$ and $t = 1000$ (f). In (d-e), lines are computed from simulations using Eq.~\eqref{eq:dynamicalinfoseries}, and shaded regions reflect analytic upper and lower bounds from Eq.~\eqref{eq:expIkULB}. In (f), $I_1$ is computed exactly using Eq.~\eqref{eq:ERW_I1}, while $I_\mathrm{tot}$ and $I_{>1} = I_\mathrm{tot}-I_1$ are approximated using Eq.~\eqref{eq:Itot_ERW} (solid lines) and bounded exactly using Eqs.~\eqref{eq:dynamicalinfoapprox} and \eqref{eq:ItotUB} (shaded regions). Note that bounds are tight enough to be nearly indistinguishable from numerical calculations. In (a-e), we set $\alpha=0.9$.}
    \label{fig:fig3}
\end{figure*}

\subsection{Poisson history dependence}

We begin with dependencies defined by a shifted Poisson distribution,
\begin{equation}
\rho(k) = \frac{\lambda^{k-1}e^{-\lambda}}{(k-1)!}.
\end{equation}
The single parameter $\lambda > 0$ determines the location of the peak of the distribution, which is given by $k = \lfloor \lambda+1\rfloor$ for non-integer $\lambda$ (and $k = \lambda$ for integer $\lambda$). For large $k$, the distribution decays faster than exponentially, severely limiting the strength of long-range dependencies [Fig.~\ref{fig:fig3}(a)].

The generating function for the dependence distribution is $\tilde{\rho}(z) = e^{-\lambda}ze^{\lambda z}$. Eq.~\eqref{eq:Cz} then gives the autocorrelation generating function,
\begin{equation}
\tilde{C}(z) \propto \frac{1}{1-(2\alpha-1)e^{-\lambda}ze^{\lambda z}}.
\end{equation}
As discussed above, the pole of $\tilde{C}(z)$ nearest to $z=1$ determines the asymptotic scaling of the correlations. Expanding the denominator of $\tilde{C}(z)$ around $z=1$ allows us to identify this pole,
\begin{equation}
z_c = \frac{1+\lambda(2\alpha-1)}{(1+\lambda)(2\alpha-1)}.
\end{equation}
Thus, near $z=1$ the autocorrelation generating function takes the form
\begin{equation}
\tilde{C}(z)\propto\frac{1}{1-z/z_c}.
\end{equation}
Using the residue theorem, we arrive at the large--$k$ scaling of the autocorrelations,
\begin{subequations}
\begin{align}
C(k\gg1) & \sim \lim_{z\to z_c}(z_c-z)\frac{\tilde{C}(z)}{z^{k+1}}\\
& = \left(\frac{1+\lambda(2\alpha-1)}{(1+\lambda)(2\alpha-1)}\right)^{-k}.\label{eq:poissonCk}
\end{align}
\end{subequations}
Thus in the large--$k$ limit the autocorrelations decay exponentially, which we confirm through numerical simulations of the copy process [Fig.~\ref{fig:fig3}(b)]. This shows that autocorrelations can decay exponentially even though the underlying history dependence decays faster. The dynamical information, by contrast, correctly captures both the shape of the history dependence distribution as well as the superexponential scaling [Fig.~\ref{fig:fig3}(c)].

\subsection{Exponential history dependence}

The simplest exponentially-decaying dependencies are defined by the geometric distribution [Fig.~\ref{fig:fig3}(a)],
\begin{equation}
\rho(k) = q (1-q)^{k-1},
\end{equation}
where $0<q\leq1$ is a shape parameter. In the limit $q=1$, $\rho(k)$ is a delta function at $k=1$, and the dynamics are Markovian. As $q$ decreases, the distribution becomes flatter, approaching a uniform distribution as $q\to 0$. 

\subsubsection{Autocorrelations}

For exponential history dependence, the autocorrelations can be computed exactly using the recurrence relation in Eq.~\eqref{autocorrelationrecurrence} along with the fact $C(0)=1$.  Beginning with an exponential ansatz $C(k) = AB^{-k}$, we obtain the self-consistent solution
\begin{equation}
C(k) = \gamma\left[1-2q+2\alpha q\right]^{k-1}, \,\,\, k\geq1,\label{eq:expCk}
\end{equation}
where $\gamma$ is a constant that depends on $q$ and $\alpha$ (see Appendix). We confirm this analytic prediction in simulations of the copy process [Fig.~\ref{fig:fig3}(b)]. Thus, even exponential dependencies that extend to infinite order still produce exponentially-decaying autocorrelations.

\subsubsection{Mapping to Markov process}

Intriguingly, we find that the copy process with geometric $\rho(k)$ can be mapped onto a first-order Markov process using the collective variable $Y(t)$ [Eq.~\eqref{eq:Ytdef}]. We can define the steady-state dynamics of $Y(t)$ with a simple Markovian update rule:
\begin{equation}
Y(t+1) = \begin{cases} (1-q) Y(t) + q, & \text{probability: } \frac{1}{2} + \frac{2\alpha-1}{2}Y(t),\\
(1-q) Y(t) - q, & \text{probability: } \frac{1}{2} - \frac{2\alpha-1}{2}Y(t). \end{cases}
\end{equation}
This mapping allows us to solve for the exact variance of the collective variable (see Appendix),
\begin{equation}\label{eq:geomYt2}
\left\langle Y(t)^2\right\rangle = \frac{q}{4 + 4\alpha q - 4\alpha - 3q}.
\end{equation}
This mapping to a Markov process facilitates the analytic investigation of information-theoretic quantities.

\subsubsection{Entropy and dynamical information}

We compute the Markovian conditional entropy $h_1$ analytically by constructing the distribution $p(x_t|x_{t-1})$ using the collective variable $Y(t)$. The calculation (detailed in the Appendix) yields
\begin{equation}\label{eq:geomh1}
h_1 = \ln2 - \Phi\left((2\alpha-1)q\left[1 + \frac{(2\alpha-1)(1-q)}{4 + 4\alpha q - 4\alpha - 3q}\right]\right).
\end{equation}
Similarly, we apply Eqs.~\eqref{eq:hinftybounds} and \eqref{eq:geomYt2} to bound the entropy rate from both above and below:
\begin{equation}
\begin{aligned}
 \ln2 &-  \Phi\left(2\alpha-1\right)\left[\frac{q}{4 + 4\alpha q - 4\alpha - 3q}\right]\leq h_\infty\\
 & \leq \ln2 -  \Phi\left((2\alpha-1)\sqrt{\frac{q}{4 + 4\alpha q - 4\alpha - 3q}}\right).
\end{aligned}
\end{equation}

The dynamical information $I_k$ can be computed numerically from simulations using Eq.~\eqref{eq:dynamicalinfoseries}, which shows it decays exponentially with $k$ and thus captures the true underlying history dependence [Fig.~\ref{fig:fig3}(c)]. For the total dynamical information, Eqs.~\eqref{eq:dynamicalinfoapprox} and \eqref{eq:ItotUB} provide analytic bounds,
\begin{equation}\begin{aligned}
\label{eq:expIkULB}
& \Phi\left((2\alpha-1)\sqrt{\frac{q}{4 + 4\alpha q - 4\alpha - 3q}}\right)\leq I_\mathrm{tot}\\
\leq & \Phi\left(2\alpha-1\right)\left[\frac{q}{4 + 4\alpha q - 4\alpha - 3q}\right].
\end{aligned}
\end{equation}
These bounds tightly constrain $I_\text{tot}$ and match the third-order truncation of Eq.~\eqref{eq:dynamicalinfoseries} computed from simulation data [Fig.~\ref{fig:fig3}(d)]. We find that the total dynamical information depends strongly on the shape parameter $q$. For $q\approx 1$, the geometric distribution approximates a delta function at $k=1$. Since the history dependence is entirely concentrated on a single past state, there is no uncertainty about which state will be copied, and thus the dynamical information is maximized. As $q$ decreases, the geometric distribution becomes flatter and thus dependencies are spread over a wider range of past states. This causes the total dynamical information $I_\text{tot}$ to decrease monotonically, eventually vanishing as $q \to 0$.

The Markovian information $I_1$ can also be computed analytically from the first-order conditional entropy [Eq.~\eqref{eq:geomh1}],
\begin{equation}
\label{eq:exp_copy_I1}
I_1 = \Phi\left((2\alpha-1)q\left[1 + \frac{(2\alpha-1)(1-q)}{4 + 4\alpha q - 4\alpha - 3q}\right]\right).
\end{equation}
In combination with the analytic bounds for $I_\mathrm{tot}$, this allows us to explore how the balance of Markovian and non-Markovian information shifts with $q$. For $q = 1$, the dynamics are entirely Markovian ($I_\text{tot} = I_1$), such that there is no information in non-Markovian dependencies ($I_{>1} = 0$). As $q$ decreases, the dynamical information is increasingly dominated by non-Markovian contributions. However, in the limit $q \to 0$, the total dynamical information vanishes and so too must the non-Markovian information. We therefore find a peak in $I_{>1}$ at intermediate $q\approx 0.2$, indicating a specific parameter regime where non-Markovian dependencies are maximized.

\subsection{Power-law history dependence}

For dynamics with longer-range dependencies, we consider a power-law history dependence distribution,
\begin{equation}
\rho(k) = \frac{k^{-s}}{\zeta(s)},
\end{equation}
where $s>1$ is the power-law exponent, and $\zeta(s)$ the Riemann zeta function, which normalizes the distribution. As $s$ decreases, the distribution becomes flatter, leading to longer-range history dependence [Fig.~\ref{fig:fig3}(a)]. 

\subsubsection{Autocorrelations for $s>2$}

We determine the autocorrelation scaling using the generating function for the history dependence, $\tilde{\rho}(z) = \mathrm{Li}_s(z)/\zeta(s)$, where $\mathrm{Li}_s(z)$ is the polylogarithm function. Since the distribution $\rho(k)$ has fundamentally different character for $s>2$ (where the mean $\langle k\rangle_\rho$ exists) and $s\leq 2$ (where the mean does not exist), we separate our analysis for these two cases.

For $s>2$, we begin by letting $z=1+u$ and expanding the polylogarithm for small $u$,
\begin{equation}
\mathrm{Li}_s(1+u) = \zeta(s) + u\zeta(s-1) + \mathcal{O}(u^2).
\end{equation}
Using this expansion, the pole of $\tilde{C}(z)$ nearest $z=1$ occurs when the denominator in Eq.~\eqref{eq:Cz} is zero, yielding
\begin{equation}
z_c = 1 + \frac{2(1-\alpha)}{\langle k\rangle_\rho(2\alpha-1)},
\end{equation}
where $\langle k\rangle_\rho = \zeta(s-1)/\zeta(s)$ is the mean value of $\rho(k)$. Then, near $z_c$ (and thus also near $z=1$), the autocorrelation generating function scales as
\begin{equation}
\tilde{C}(z) \propto \frac{1}{1-z/z_c}.
\end{equation}
Since the pole at $z=z_c$ is simple, we can evaluate the autocorrelations from Eq.~\eqref{eq:cauchy} using the residue theorem:
\begin{equation}
C(k)\propto \left(1 + \frac{2(1-\alpha)}{\langle k\rangle_\rho(2\alpha-1)}\right)^{-k}.\label{eq:zetaCk}
\end{equation}

Surprisingly, the autocorrelations appear to decay exponentially, and we confirm this exponential scaling in simulations [Fig.~\ref{fig:fig3}(b)]. However, this is not the complete story. The simple pole at $z_c$ determines the scaling up to a finite threshold $k^*$ that depends on $s$ and $\alpha$. In the asymptotic limit $k\to\infty$, the branch point of the polylogarithm at $z=1$ eventually dominates. From standard generating function transfer theorems~\cite{flajolet1990singularity, flajolet2003singular}, the large-$k$ scaling will be $C(k)\sim k^{-s}$ for noninteger $s$, and $C(k)\propto k^{-s}\ln s$ for integer $s$ (see Appendix). Thus the power-law autocorrelations expected from intuition are eventually recovered at sufficiently large times.

Thus, for power-law history dependence with exponent $s>2$, the autocorrelations exhibit two qualitatively distinct regimes. An exponential regime for finite $k$, which transitions to power-law scaling for large $k$. While the exponential scaling regime is necessarily finite, the crossover point can be made arbitrarily large by tuning $s$ and $\alpha$. We show in the Appendix that the crossover $k^*$ scales approximately as
\begin{equation}
k^*\sim -\frac{s}{\ln z_c}W_{-1}\left(-\frac{\ln z_c}{s}\right),
\end{equation}
where $W_{-1}(z)$ is the $-1$ branch of the Lambert W function. For example, with $s=4$ and $\alpha=0.9$, the crossover occurs at  $k^*\sim88$. This explains why the autocorrelations appear to follow pure exponential scaling in Fig.~\ref{fig:fig3}(b). Thus, even for a non-Markovian process with power-law history dependence, it is nonetheless still possible to observe exponentially-decaying autocorrelations for any finite amount of data. In turn, this establishes that observing exponential autocorrelations cannot rule out the possibility of truly long-range dependencies.

\subsubsection{Autocorrelations for $s\leq 2$}

For $s\leq 2$, the dependence distribution becomes more heavy-tailed, and $\rho(k)$ no longer has a finite mean. In this case, $\tilde{C}(z)$ does not have a simple pole arbitrarily close to $z=1$. We can nonetheless obtain the asymptotic scaling of the autocorrelations using transfer theorem arguments~\cite{flajolet1990singularity,flajolet2003singular} (see Appendix), which yield
\begin{equation}
C(k)\sim\begin{cases} k^{-2}\ln k, & s=2,\\
k^{-s}, & 1<s<2.
\end{cases}
\end{equation}
Thus, for sufficiently strong power-law history dependence $s \le 2$, we find that the correlations do indeed decay as a power law for all $k$. We confirm this power-law scaling in simulations [Fig.~\ref{fig:fig3}(b)].

\subsubsection{Entropy and dynamical information}

Unlike exponential dependencies, for power-law dependencies it is not possible to map the copy model onto a Markov process. As such, we cannot analytically determine the statistics of the collective variable, nor analytically approximate information-theoretic quantities. We can nonetheless still compute the dynamical information directly from simulation data using a truncation of the infinite series~\eqref{eq:dynamicalinfoseries}. We find in general that the dynamical information scales as a power law with $k$, as required by Eq.~\eqref{eq:dyninfobound}). Thus the dynamical information correctly captures the underlying power-law dependencies, even for $s<2$ when the autocorrelations do not [Fig.~\ref{fig:fig3}(c)].

For each power-law exponent $s$, the total dynamical information $I_\text{tot}$ arises from a different mixture of Markovian and non-Markovian contributions [Fig.~\ref{fig:fig3}(e)]. As $s$ increases and the dependencies become flatter, the total dynamical information decreases while becoming increasingly dominated by non-Markovian contributions. As a result, we observe a peak in the non-Markovian information $I_{>2}$ at $s\approx2$. Notably, this peak roughly corresponds to the transition from finite average dependence length $\langle k\rangle_\rho$ to infinite length.

\subsection{Elephant random walk}

Taking a final step in the direction of longer-range dependencies leads to the well-studied Elephant Random Walk (ERW)~\cite{schutz2004elephants}, in which $\rho(k)$ is uniform over the entire history of the process [Fig.~\ref{fig:fig3}(a)]. For the ERW, the conditional distribution for the dynamics must be defined with explicit time-dependence. We initialize the process at $t=0$ with a random state from $\{-1,1\}$, so that the conditional distribution for $t\geq 1$ is given by
\begin{equation}
p(x_t|x_{t-1},\hdots ,x_0) = \frac{1}{2} + \frac{2\alpha-1}{2t}\sum_{k=0}^{t-1} x_k x_{t}.
\end{equation}

\subsubsection{Autocorrelations}

Autocorrelations for the ERW have been calculated exactly for all $t$ (for details see Ref.~\cite{coletti2017central}). In the long-time limit, for $t,k\to\infty$ with fixed $k/t$, the autocorrelations scale as~\cite{coletti2017central}
\begin{equation}
C(k,t) \sim \left(1 + \frac{k}{t}\right)^{-2(1-\alpha)}F(t,\alpha),
\end{equation}
where
\begin{equation}
F(t,\alpha)\sim\begin{cases}
\frac{2(2\alpha-1)(1-\alpha)}{3-4\alpha}t^{-1}, & \alpha<3/4,\\
\frac{\ln t}{4t}, & \alpha=3/4,\\
\frac{(2\alpha-1)^2}{(4\alpha-3)\Gamma(4\alpha-2)}t^{-4(1-\alpha)}, & \alpha>3/4.
\end{cases}
\end{equation}
Thus, for all values of $\alpha$, the ERW produces power-law autocorrelations.

\subsubsection{Collective variable}

Analysis of the ERW is simplified significantly by studying the collective variable $Y(t) = \frac{1}{t}\sum_{k=0}^{t-1}x_k$. It has been shown previously that the first three moments can be calculated exactly in the large-$t$ limit~\cite{coletti2017central}, yielding $\langle Y(t)\rangle = \langle Y(t)^3\rangle = 0 $ and
\begin{equation}\label{eq:ERW_Yt}
\langle Y(t)^2\rangle = \begin{cases}
\frac{1}{(3-4\alpha)t}, & \alpha<3/4,\\
\frac{\ln t}{t}, & \alpha=3/4,\\
\frac{t^{4\alpha-4}}{(4\alpha-3)\Gamma(4\alpha-1)}, & \alpha>3/4.
\end{cases}
\end{equation}
The process undergoes a phase transition at $\alpha=0.75$. For $\alpha\leq0.75$, the collective variable $Y(t)$ is Gaussian-distributed in the long-time limit, while for $\alpha>0.75$, the collective variable converges to a non-Gaussian distribution~\cite{coletti2017central}, which has not been fully characterized in the literature.

Just as for exponentially-decaying history dependence, the ERW can be mapped to a first-order Markov process. The dynamics of the collective variable $Y(t)$ are defined by a simple Markovian update rule:
\begin{equation}
Y(t+1) = \begin{cases} \frac{t}{t+1}Y(t) + \frac{1}{t+1}, & \text{probability: } \frac{1}{2} + \frac{2\alpha-1}{2}Y(t),\\
\frac{t}{t+1} Y(t) - \frac{1}{t+1}, & \text{probability: } \frac{1}{2} - \frac{2\alpha-1}{2}Y(t). \end{cases}
\end{equation}
Thus, the fact that a non-Markovian process can be mapped to a Markov process does not imply exponentially-decaying autocorrelations; these features appear to be orthogonal.

\subsubsection{Entropy and dynamical information}

Taking advantage of the Markov process mapping as well as the known Gaussian statistics of $Y(t)$ for $\alpha\leq3/4$~\cite{coletti2017central}, we derive a recursive formula for the conditional entropy $h_k$ to arbitrarily large $k$ (see Appendix). Furthermore, combining Eq.~\eqref{eq:ERW_Yt} with Eq.~\eqref{eq:h_kseries} (truncated to $n=1$) yields an approximation for the entropy rate, 
\begin{equation}
\label{eq:ERW_hinfty}
h_\infty \approx \ln2 - \frac{1}{2}(2\alpha-1)^2 \times \begin{cases}
\frac{1}{(3-4\alpha)t}, & \alpha<3/4,\\
\frac{\ln t}{t}, & \alpha=3/4,\\
\frac{t^{4\alpha-4}}{(4\alpha-3)\Gamma(4\alpha-1)}, & \alpha>3/4.
\end{cases}
\end{equation}

The recursive formula for the conditional entropy $h_k$ [Eq.~\eqref{eq:h_k_recursion}] enables us to numerically compute the dynamical information $I_k$, albeit at a computational cost that grows exponentially with $k$. In practice, we are able to compute $I_k$ up to $k=20$, and we find that it decays sub-linearly with $k$ [Fig.~\ref{fig:fig3}(c)]. Using Eq.~\eqref{eq:ERW_hinfty}, we have the following analytic approximation for the total dynamical information,
\begin{equation}
\label{eq:Itot_ERW}
I_\mathrm{tot} \approx \frac{1}{2}(2\alpha-1)^2 \times \begin{cases}
\frac{1}{(3-4\alpha)t}, & \alpha<3/4,\\
\frac{\ln t}{t}, & \alpha=3/4,\\
\frac{t^{4\alpha-4}}{(4\alpha-3)\Gamma(4\alpha-1)}, & \alpha>3/4.
\end{cases}
\end{equation}
This dynamical information can also be bounded from both above and below by combining Eqs.~\eqref{eq:dynamicalinfoapprox} and \eqref{eq:ItotUB} with the asymptotic variance of the collective variable in Eq.~\eqref{eq:ERW_Yt}.
For all $\alpha$, the dynamical information decreases monotonically as $t\to\infty$. This can be understood intuitively by noting that as $t$ increases, the history dependence becomes spread more thinly across the entire history, ultimately leading to complete uncertainty about the next state even when the entire history is known. Counterintuitively, $I_\text{tot}$ is non-monotonic as a function of the copying fidelity $\alpha$ [Fig.~\ref{fig:fig3}(f)]. Although the total dynamical information generally increases with $\alpha$, it exhibits both a cusp and global maximum at the critical point $\alpha = 0.75$.

Similarly, we observe cusps and global maximuma in both the Markovian and non-Markovian dynamical information at the critical point $\alpha=0.75$ [Fig.~\ref{fig:fig3}(f)]. We likewise find that the proportion of Markovian information as a function of total dynamical information is maximized at $\alpha=0.75$. Finally, we note that the Markovian information $I_1$ can be computed exactly in the long-time limit (see Appendix), yielding
\begin{equation}\label{eq:ERW_I1}
I_1 = \Phi\left(\frac{1}{t}(2\alpha-1) \left[1 + (2\alpha-1)(t-1)\left\langle Y(t-1)^2\right\rangle\right]\right).
\end{equation}

Together, our results for infinite-order non-Markovian dynamics reveal the complex relationships between dependencies, correlations, and information. Notably, exponential autocorrelations arise from a range of qualitatively distinct underlying dependencies [Fig.~\ref{fig:fig3}(b)], while dynamical information consistently reflects the true history dependence. Additionally, we find that the amount of non-Markovian information $I_{>1}$ does not simply increase as historical dependencies become stronger [Fig.~\ref{fig:fig3}(d-f)].

\section{Discussion}

Non-Markovian dynamics arise across all fields that rely on coarse-grained descriptions of the laws of physics. To better understand these processes, tractable models are needed to guide our intuition. In this paper, we define and analyze a non-Markovian process in which the current state attempts to copy the state $k$ steps in the past with probably $\rho(k)$. By tuning the distribution $\rho(k)$, this copy model is flexible enough to capture a wide variety of historical dependence structures, generalizing the well-studied Elephant Random Walk~\cite{schutz2004elephants}. This model is analytically tractable, allowing us to obtain both general results for arbitrary $\rho(k)$ (Sec.~\ref{sec:generalresults}) and precise results for specific choices of $\rho(k)$ (Secs.~\ref{sec:finiteorder} and \ref{sec:infiniteorder}). This tractability opens the door for investigations into the interplay between history dependence, autocorrelations, and information-theoretic quantities like entropy and dynamical information~\cite{leighton2025main}.

As perhaps our most striking result, we find that autocorrelations can fundamentally fail to reflect true history dependence in non-Markovian processes. Observing exponentially-decaying autocorrelations in finite data is consistent with many underlying dependencies: Markovian, finite-order non-Markovian, and infinite-order non-Markovian with superexponential, exponential, and even power-law decay. This finding is at odds with the traditional intuition that non-Markovian processes with long-range dependencies generate power-law autocorrelations, and suggests that care must be taken when drawing conclusions about dependencies from autocorrelations. By contrast, the dynamical information (proposed in the companion paper~\cite{leighton2025main}) more accurately captures underlying history dependencies.

We have only scratched the surface of the rich behaviors that this non-Markovian copy process can produce. In particular, further analytic study of the copy process with power-law-distributed $\rho(k)$ would be of significant interest, particularly the transition in autocorrelation scaling at $s=2$. The non-Markovian copy process may also provide a minimal model for the transformer architecture in machine learning, with the history dependence $\rho(k)$ serving a function analogous to the attention mechanism~\cite{vaswani2017attention}. More generally, the ability to tune $\rho(k)$ directly enables future investigations into many classes of historical dependencies not considered here.

\emph{Acknowledgements}---This work was supported in part by Mossman and NSERC Postdoctoral Fellowships (M.P.L.), and by support from the Department of Physics and the Quantitative Biology Institute at Yale University (C.W.L).
\bibliography{main}

\begin{thebibliography}{37}%
\makeatletter
\providecommand \@ifxundefined [1]{%
 \@ifx{#1\undefined}
}%
\providecommand \@ifnum [1]{%
 \ifnum #1\expandafter \@firstoftwo
 \else \expandafter \@secondoftwo
 \fi
}%
\providecommand \@ifx [1]{%
 \ifx #1\expandafter \@firstoftwo
 \else \expandafter \@secondoftwo
 \fi
}%
\providecommand \natexlab [1]{#1}%
\providecommand \enquote  [1]{``#1''}%
\providecommand \bibnamefont  [1]{#1}%
\providecommand \bibfnamefont [1]{#1}%
\providecommand \citenamefont [1]{#1}%
\providecommand \href@noop [0]{\@secondoftwo}%
\providecommand \href [0]{\begingroup \@sanitize@url \@href}%
\providecommand \@href[1]{\@@startlink{#1}\@@href}%
\providecommand \@@href[1]{\endgroup#1\@@endlink}%
\providecommand \@sanitize@url [0]{\catcode `\\12\catcode `\$12\catcode
  `\&12\catcode `\#12\catcode `\^12\catcode `\_12\catcode `\%12\relax}%
\providecommand \@@startlink[1]{}%
\providecommand \@@endlink[0]{}%
\providecommand \url  [0]{\begingroup\@sanitize@url \@url }%
\providecommand \@url [1]{\endgroup\@href {#1}{\urlprefix }}%
\providecommand \urlprefix  [0]{URL }%
\providecommand \Eprint [0]{\href }%
\providecommand \doibase [0]{https://doi.org/}%
\providecommand \selectlanguage [0]{\@gobble}%
\providecommand \bibinfo  [0]{\@secondoftwo}%
\providecommand \bibfield  [0]{\@secondoftwo}%
\providecommand \translation [1]{[#1]}%
\providecommand \BibitemOpen [0]{}%
\providecommand \bibitemStop [0]{}%
\providecommand \bibitemNoStop [0]{.\EOS\space}%
\providecommand \EOS [0]{\spacefactor3000\relax}%
\providecommand \BibitemShut  [1]{\csname bibitem#1\endcsname}%
\let\auto@bib@innerbib\@empty
\bibitem [{\citenamefont {Vroylandt}\ \emph {et~al.}(2022)\citenamefont
  {Vroylandt}, \citenamefont {Gouden{\`e}ge}, \citenamefont {Monmarch{\'e}},
  \citenamefont {Pietrucci},\ and\ \citenamefont
  {Rotenberg}}]{vroylandt2022likelihood}%
  \BibitemOpen
  \bibfield  {author} {\bibinfo {author} {\bibfnamefont {H.}~\bibnamefont
  {Vroylandt}}, \bibinfo {author} {\bibfnamefont {L.}~\bibnamefont
  {Gouden{\`e}ge}}, \bibinfo {author} {\bibfnamefont {P.}~\bibnamefont
  {Monmarch{\'e}}}, \bibinfo {author} {\bibfnamefont {F.}~\bibnamefont
  {Pietrucci}},\ and\ \bibinfo {author} {\bibfnamefont {B.}~\bibnamefont
  {Rotenberg}},\ }\bibfield  {title} {\bibinfo {title} {Likelihood-based
  non-{M}arkovian models from molecular dynamics},\ }\href@noop {} {\bibfield
  {journal} {\bibinfo  {journal} {Proceedings of the National Academy of
  Sciences}\ }\textbf {\bibinfo {volume} {119}},\ \bibinfo {pages}
  {e2117586119} (\bibinfo {year} {2022})}\BibitemShut {NoStop}%
\bibitem [{\citenamefont {D’Urso}\ and\ \citenamefont
  {Brickner}(2014)}]{d2014mechanisms}%
  \BibitemOpen
  \bibfield  {author} {\bibinfo {author} {\bibfnamefont {A.}~\bibnamefont
  {D’Urso}}\ and\ \bibinfo {author} {\bibfnamefont {J.~H.}\ \bibnamefont
  {Brickner}},\ }\bibfield  {title} {\bibinfo {title} {Mechanisms of epigenetic
  memory},\ }\href@noop {} {\bibfield  {journal} {\bibinfo  {journal} {Trends
  in Genetics}\ }\textbf {\bibinfo {volume} {30}},\ \bibinfo {pages} {230}
  (\bibinfo {year} {2014})}\BibitemShut {NoStop}%
\bibitem [{\citenamefont {Zeraati}\ \emph {et~al.}(2023)\citenamefont
  {Zeraati}, \citenamefont {Shi}, \citenamefont {Steinmetz}, \citenamefont
  {Gieselmann}, \citenamefont {Thiele}, \citenamefont {Moore}, \citenamefont
  {Levina},\ and\ \citenamefont {Engel}}]{zeraati2023intrinsic}%
  \BibitemOpen
  \bibfield  {author} {\bibinfo {author} {\bibfnamefont {R.}~\bibnamefont
  {Zeraati}}, \bibinfo {author} {\bibfnamefont {Y.-L.}\ \bibnamefont {Shi}},
  \bibinfo {author} {\bibfnamefont {N.~A.}\ \bibnamefont {Steinmetz}}, \bibinfo
  {author} {\bibfnamefont {M.~A.}\ \bibnamefont {Gieselmann}}, \bibinfo
  {author} {\bibfnamefont {A.}~\bibnamefont {Thiele}}, \bibinfo {author}
  {\bibfnamefont {T.}~\bibnamefont {Moore}}, \bibinfo {author} {\bibfnamefont
  {A.}~\bibnamefont {Levina}},\ and\ \bibinfo {author} {\bibfnamefont {T.~A.}\
  \bibnamefont {Engel}},\ }\bibfield  {title} {\bibinfo {title} {Intrinsic
  timescales in the visual cortex change with selective attention and reflect
  spatial connectivity},\ }\href@noop {} {\bibfield  {journal} {\bibinfo
  {journal} {Nature Communications}\ }\textbf {\bibinfo {volume} {14}},\
  \bibinfo {pages} {1858} (\bibinfo {year} {2023})}\BibitemShut {NoStop}%
\bibitem [{\citenamefont {Cavanagh}\ \emph {et~al.}(2020)\citenamefont
  {Cavanagh}, \citenamefont {Hunt},\ and\ \citenamefont
  {Kennerley}}]{cavanagh2020diversity}%
  \BibitemOpen
  \bibfield  {author} {\bibinfo {author} {\bibfnamefont {S.~E.}\ \bibnamefont
  {Cavanagh}}, \bibinfo {author} {\bibfnamefont {L.~T.}\ \bibnamefont {Hunt}},\
  and\ \bibinfo {author} {\bibfnamefont {S.~W.}\ \bibnamefont {Kennerley}},\
  }\bibfield  {title} {\bibinfo {title} {A diversity of intrinsic timescales
  underlie neural computations},\ }\href@noop {} {\bibfield  {journal}
  {\bibinfo  {journal} {Frontiers in neural circuits}\ }\textbf {\bibinfo
  {volume} {14}},\ \bibinfo {pages} {615626} (\bibinfo {year}
  {2020})}\BibitemShut {NoStop}%
\bibitem [{\citenamefont {ElGamel}\ \emph {et~al.}(2023)\citenamefont
  {ElGamel}, \citenamefont {Vashistha}, \citenamefont {Salman},\ and\
  \citenamefont {Mugler}}]{elgamel2023multigenerational}%
  \BibitemOpen
  \bibfield  {author} {\bibinfo {author} {\bibfnamefont {M.}~\bibnamefont
  {ElGamel}}, \bibinfo {author} {\bibfnamefont {H.}~\bibnamefont {Vashistha}},
  \bibinfo {author} {\bibfnamefont {H.}~\bibnamefont {Salman}},\ and\ \bibinfo
  {author} {\bibfnamefont {A.}~\bibnamefont {Mugler}},\ }\bibfield  {title}
  {\bibinfo {title} {Multigenerational memory in bacterial size control},\
  }\href@noop {} {\bibfield  {journal} {\bibinfo  {journal} {Physical Review
  E}\ }\textbf {\bibinfo {volume} {108}},\ \bibinfo {pages} {L032401} (\bibinfo
  {year} {2023})}\BibitemShut {NoStop}%
\bibitem [{\citenamefont {Alba}\ \emph {et~al.}(2020)\citenamefont {Alba},
  \citenamefont {Berman}, \citenamefont {Bialek},\ and\ \citenamefont
  {Shaevitz}}]{alba2020exploring}%
  \BibitemOpen
  \bibfield  {author} {\bibinfo {author} {\bibfnamefont {V.}~\bibnamefont
  {Alba}}, \bibinfo {author} {\bibfnamefont {G.~J.}\ \bibnamefont {Berman}},
  \bibinfo {author} {\bibfnamefont {W.}~\bibnamefont {Bialek}},\ and\ \bibinfo
  {author} {\bibfnamefont {J.~W.}\ \bibnamefont {Shaevitz}},\ }\bibfield
  {title} {\bibinfo {title} {Exploring a strongly non-{M}arkovian animal
  behavior},\ }\href@noop {} {\bibfield  {journal} {\bibinfo  {journal} {arXiv
  preprint arXiv:2012.15681}\ } (\bibinfo {year} {2020})}\BibitemShut {NoStop}%
\bibitem [{\citenamefont {Bialek}\ and\ \citenamefont
  {Shaevitz}(2024)}]{bialek2024long}%
  \BibitemOpen
  \bibfield  {author} {\bibinfo {author} {\bibfnamefont {W.}~\bibnamefont
  {Bialek}}\ and\ \bibinfo {author} {\bibfnamefont {J.~W.}\ \bibnamefont
  {Shaevitz}},\ }\bibfield  {title} {\bibinfo {title} {Long timescales,
  individual differences, and scale invariance in animal behavior},\
  }\href@noop {} {\bibfield  {journal} {\bibinfo  {journal} {Physical Review
  Letters}\ }\textbf {\bibinfo {volume} {132}},\ \bibinfo {pages} {048401}
  (\bibinfo {year} {2024})}\BibitemShut {NoStop}%
\bibitem [{\citenamefont {Barabasi}(2005)}]{barabasi2005origin}%
  \BibitemOpen
  \bibfield  {author} {\bibinfo {author} {\bibfnamefont {A.-L.}\ \bibnamefont
  {Barabasi}},\ }\bibfield  {title} {\bibinfo {title} {The origin of bursts and
  heavy tails in human dynamics},\ }\href@noop {} {\bibfield  {journal}
  {\bibinfo  {journal} {Nature}\ }\textbf {\bibinfo {volume} {435}},\ \bibinfo
  {pages} {207} (\bibinfo {year} {2005})}\BibitemShut {NoStop}%
\bibitem [{\citenamefont {Shannon}(1948)}]{shannon1948mathematical}%
  \BibitemOpen
  \bibfield  {author} {\bibinfo {author} {\bibfnamefont {C.~E.}\ \bibnamefont
  {Shannon}},\ }\bibfield  {title} {\bibinfo {title} {A mathematical theory of
  communication},\ }\href@noop {} {\bibfield  {journal} {\bibinfo  {journal}
  {The Bell System Technical Journal}\ }\textbf {\bibinfo {volume} {27}},\
  \bibinfo {pages} {379} (\bibinfo {year} {1948})}\BibitemShut {NoStop}%
\bibitem [{\citenamefont {Wieczynski}\ and\ \citenamefont
  {Debowski}(2025)}]{wieczynski2025long}%
  \BibitemOpen
  \bibfield  {author} {\bibinfo {author} {\bibfnamefont {P.}~\bibnamefont
  {Wieczynski}}\ and\ \bibinfo {author} {\bibfnamefont {L.}~\bibnamefont
  {Debowski}},\ }\bibfield  {title} {\bibinfo {title} {Long-range dependence in
  word time series: The cosine correlation of embeddings},\ }\href@noop {}
  {\bibfield  {journal} {\bibinfo  {journal} {Entropy}\ }\textbf {\bibinfo
  {volume} {27}},\ \bibinfo {pages} {613} (\bibinfo {year} {2025})}\BibitemShut
  {NoStop}%
\bibitem [{\citenamefont {Breuer}\ \emph {et~al.}(2016)\citenamefont {Breuer},
  \citenamefont {Laine}, \citenamefont {Piilo},\ and\ \citenamefont
  {Vacchini}}]{breuer2016colloquium}%
  \BibitemOpen
  \bibfield  {author} {\bibinfo {author} {\bibfnamefont {H.-P.}\ \bibnamefont
  {Breuer}}, \bibinfo {author} {\bibfnamefont {E.-M.}\ \bibnamefont {Laine}},
  \bibinfo {author} {\bibfnamefont {J.}~\bibnamefont {Piilo}},\ and\ \bibinfo
  {author} {\bibfnamefont {B.}~\bibnamefont {Vacchini}},\ }\bibfield  {title}
  {\bibinfo {title} {Colloquium: Non-markovian dynamics in open quantum
  systems},\ }\href@noop {} {\bibfield  {journal} {\bibinfo  {journal} {Reviews
  of Modern Physics}\ }\textbf {\bibinfo {volume} {88}},\ \bibinfo {pages}
  {021002} (\bibinfo {year} {2016})}\BibitemShut {NoStop}%
\bibitem [{\citenamefont {De~Vega}\ and\ \citenamefont
  {Alonso}(2017)}]{de2017dynamics}%
  \BibitemOpen
  \bibfield  {author} {\bibinfo {author} {\bibfnamefont {I.}~\bibnamefont
  {De~Vega}}\ and\ \bibinfo {author} {\bibfnamefont {D.}~\bibnamefont
  {Alonso}},\ }\bibfield  {title} {\bibinfo {title} {Dynamics of non-markovian
  open quantum systems},\ }\href@noop {} {\bibfield  {journal} {\bibinfo
  {journal} {Reviews of Modern Physics}\ }\textbf {\bibinfo {volume} {89}},\
  \bibinfo {pages} {015001} (\bibinfo {year} {2017})}\BibitemShut {NoStop}%
\bibitem [{\citenamefont {Bogun{\'a}}\ \emph {et~al.}(2013)\citenamefont
  {Bogun{\'a}}, \citenamefont {Lafuerza}, \citenamefont {Toral},\ and\
  \citenamefont {Serrano}}]{boguna2013simulating}%
  \BibitemOpen
  \bibfield  {author} {\bibinfo {author} {\bibfnamefont {M.}~\bibnamefont
  {Bogun{\'a}}}, \bibinfo {author} {\bibfnamefont {L.~F.}\ \bibnamefont
  {Lafuerza}}, \bibinfo {author} {\bibfnamefont {R.}~\bibnamefont {Toral}},\
  and\ \bibinfo {author} {\bibfnamefont {M.}~\bibnamefont {Serrano}},\
  }\bibfield  {title} {\bibinfo {title} {Simulating non-markovian stochastic
  processes},\ }\href@noop {} {\bibfield  {journal} {\bibinfo  {journal} {arXiv
  preprint arXiv:1310.0926}\ } (\bibinfo {year} {2013})}\BibitemShut {NoStop}%
\bibitem [{\citenamefont {da~Silva}\ \emph {et~al.}(2015)\citenamefont
  {da~Silva}, \citenamefont {Viswanathan},\ and\ \citenamefont
  {Cressoni}}]{da2015two}%
  \BibitemOpen
  \bibfield  {author} {\bibinfo {author} {\bibfnamefont {M.}~\bibnamefont
  {da~Silva}}, \bibinfo {author} {\bibfnamefont {G.}~\bibnamefont
  {Viswanathan}},\ and\ \bibinfo {author} {\bibfnamefont {J.}~\bibnamefont
  {Cressoni}},\ }\bibfield  {title} {\bibinfo {title} {A two-dimensional
  non-markovian random walk leading to anomalous diffusion},\ }\href@noop {}
  {\bibfield  {journal} {\bibinfo  {journal} {Physica A: Statistical Mechanics
  and its Applications}\ }\textbf {\bibinfo {volume} {421}},\ \bibinfo {pages}
  {522} (\bibinfo {year} {2015})}\BibitemShut {NoStop}%
\bibitem [{\citenamefont {Rabiner}(2002)}]{rabiner2002tutorial}%
  \BibitemOpen
  \bibfield  {author} {\bibinfo {author} {\bibfnamefont {L.~R.}\ \bibnamefont
  {Rabiner}},\ }\bibfield  {title} {\bibinfo {title} {A tutorial on hidden
  {M}arkov models and selected applications in speech recognition},\
  }\href@noop {} {\bibfield  {journal} {\bibinfo  {journal} {Proceedings of the
  IEEE}\ }\textbf {\bibinfo {volume} {77}},\ \bibinfo {pages} {257} (\bibinfo
  {year} {2002})}\BibitemShut {NoStop}%
\bibitem [{\citenamefont {Zucchini}\ and\ \citenamefont
  {MacDonald}(2009)}]{zucchini2009hidden}%
  \BibitemOpen
  \bibfield  {author} {\bibinfo {author} {\bibfnamefont {W.}~\bibnamefont
  {Zucchini}}\ and\ \bibinfo {author} {\bibfnamefont {I.~L.}\ \bibnamefont
  {MacDonald}},\ }\href@noop {} {\emph {\bibinfo {title} {Hidden {M}arkov
  models for time series: an introduction using {R}}}}\ (\bibinfo  {publisher}
  {Chapman and Hall/CRC},\ \bibinfo {year} {2009})\BibitemShut {NoStop}%
\bibitem [{\citenamefont {Agon}\ \emph {et~al.}(2018)\citenamefont {Agon},
  \citenamefont {Balasubramanian}, \citenamefont {Kasko},\ and\ \citenamefont
  {Lawrence}}]{agon2018coarse}%
  \BibitemOpen
  \bibfield  {author} {\bibinfo {author} {\bibfnamefont {C.}~\bibnamefont
  {Agon}}, \bibinfo {author} {\bibfnamefont {V.}~\bibnamefont
  {Balasubramanian}}, \bibinfo {author} {\bibfnamefont {S.}~\bibnamefont
  {Kasko}},\ and\ \bibinfo {author} {\bibfnamefont {A.}~\bibnamefont
  {Lawrence}},\ }\bibfield  {title} {\bibinfo {title} {Coarse grained quantum
  dynamics},\ }\href@noop {} {\bibfield  {journal} {\bibinfo  {journal}
  {Physical Review D}\ }\textbf {\bibinfo {volume} {98}},\ \bibinfo {pages}
  {025019} (\bibinfo {year} {2018})}\BibitemShut {NoStop}%
\bibitem [{\citenamefont {Strasberg}\ and\ \citenamefont
  {Esposito}(2019)}]{strasberg2019non}%
  \BibitemOpen
  \bibfield  {author} {\bibinfo {author} {\bibfnamefont {P.}~\bibnamefont
  {Strasberg}}\ and\ \bibinfo {author} {\bibfnamefont {M.}~\bibnamefont
  {Esposito}},\ }\bibfield  {title} {\bibinfo {title} {Non-{M}arkovianity and
  negative entropy production rates},\ }\href@noop {} {\bibfield  {journal}
  {\bibinfo  {journal} {Physical Review E}\ }\textbf {\bibinfo {volume} {99}},\
  \bibinfo {pages} {012120} (\bibinfo {year} {2019})}\BibitemShut {NoStop}%
\bibitem [{\citenamefont {Schwarz}\ \emph {et~al.}(2024)\citenamefont
  {Schwarz}, \citenamefont {Kolomeisky},\ and\ \citenamefont
  {Godec}}]{schwarz2024mind}%
  \BibitemOpen
  \bibfield  {author} {\bibinfo {author} {\bibfnamefont {T.}~\bibnamefont
  {Schwarz}}, \bibinfo {author} {\bibfnamefont {A.~B.}\ \bibnamefont
  {Kolomeisky}},\ and\ \bibinfo {author} {\bibfnamefont {A.}~\bibnamefont
  {Godec}},\ }\bibfield  {title} {\bibinfo {title} {Mind the memory: Consistent
  time reversal removes artefactual scaling of energy dissipation rate and
  provides more accurate and reliable thermodynamic inference},\ }\href@noop {}
  {\bibfield  {journal} {\bibinfo  {journal} {arXiv preprint arXiv:2410.11819}\
  } (\bibinfo {year} {2024})}\BibitemShut {NoStop}%
\bibitem [{\citenamefont {Mochihashi}\ and\ \citenamefont
  {Sumita}(2007)}]{mochihashi2007infinite}%
  \BibitemOpen
  \bibfield  {author} {\bibinfo {author} {\bibfnamefont {D.}~\bibnamefont
  {Mochihashi}}\ and\ \bibinfo {author} {\bibfnamefont {E.}~\bibnamefont
  {Sumita}},\ }\bibfield  {title} {\bibinfo {title} {The infinite markov
  model},\ }\href@noop {} {\bibfield  {journal} {\bibinfo  {journal} {Advances
  in neural information processing systems}\ }\textbf {\bibinfo {volume} {20}}
  (\bibinfo {year} {2007})}\BibitemShut {NoStop}%
\bibitem [{\citenamefont {Marzen}\ and\ \citenamefont
  {Crutchfield}(2016)}]{marzen2016predictive}%
  \BibitemOpen
  \bibfield  {author} {\bibinfo {author} {\bibfnamefont {S.~E.}\ \bibnamefont
  {Marzen}}\ and\ \bibinfo {author} {\bibfnamefont {J.~P.}\ \bibnamefont
  {Crutchfield}},\ }\bibfield  {title} {\bibinfo {title} {Predictive
  rate-distortion for infinite-order markov processes},\ }\href@noop {}
  {\bibfield  {journal} {\bibinfo  {journal} {Journal of Statistical Physics}\
  }\textbf {\bibinfo {volume} {163}},\ \bibinfo {pages} {1312} (\bibinfo {year}
  {2016})}\BibitemShut {NoStop}%
\bibitem [{\citenamefont {Sch{\"u}tz}\ and\ \citenamefont
  {Trimper}(2004)}]{schutz2004elephants}%
  \BibitemOpen
  \bibfield  {author} {\bibinfo {author} {\bibfnamefont {G.~M.}\ \bibnamefont
  {Sch{\"u}tz}}\ and\ \bibinfo {author} {\bibfnamefont {S.}~\bibnamefont
  {Trimper}},\ }\bibfield  {title} {\bibinfo {title} {Elephants can always
  remember: Exact long-range memory effects in a non-{M}arkovian random walk},\
  }\href@noop {} {\bibfield  {journal} {\bibinfo  {journal} {Physical Review
  E}\ }\textbf {\bibinfo {volume} {70}},\ \bibinfo {pages} {045101} (\bibinfo
  {year} {2004})}\BibitemShut {NoStop}%
\bibitem [{\citenamefont {Georgiou}\ \emph {et~al.}(2015)\citenamefont
  {Georgiou}, \citenamefont {Kiss},\ and\ \citenamefont
  {Scalas}}]{georgiou2015solvable}%
  \BibitemOpen
  \bibfield  {author} {\bibinfo {author} {\bibfnamefont {N.}~\bibnamefont
  {Georgiou}}, \bibinfo {author} {\bibfnamefont {I.~Z.}\ \bibnamefont {Kiss}},\
  and\ \bibinfo {author} {\bibfnamefont {E.}~\bibnamefont {Scalas}},\
  }\bibfield  {title} {\bibinfo {title} {Solvable non-markovian dynamic
  network},\ }\href@noop {} {\bibfield  {journal} {\bibinfo  {journal}
  {Physical Review E}\ }\textbf {\bibinfo {volume} {92}},\ \bibinfo {pages}
  {042801} (\bibinfo {year} {2015})}\BibitemShut {NoStop}%
\bibitem [{\citenamefont {Saha}(2022)}]{saha2022random}%
  \BibitemOpen
  \bibfield  {author} {\bibinfo {author} {\bibfnamefont {S.}~\bibnamefont
  {Saha}},\ }\bibfield  {title} {\bibinfo {title} {Random walk with multiple
  memory channels},\ }\href@noop {} {\bibfield  {journal} {\bibinfo  {journal}
  {Physical Review E}\ }\textbf {\bibinfo {volume} {106}},\ \bibinfo {pages}
  {L062105} (\bibinfo {year} {2022})}\BibitemShut {NoStop}%
\bibitem [{\citenamefont {Boyer}\ \emph {et~al.}(2025)\citenamefont {Boyer}
  \emph {et~al.}}]{boyer2025intermittent}%
  \BibitemOpen
  \bibfield  {author} {\bibinfo {author} {\bibfnamefont {D.}~\bibnamefont
  {Boyer}} \emph {et~al.},\ }\bibfield  {title} {\bibinfo {title} {Intermittent
  localization and fast spatial learning by non-{M}arkov random walks with
  decaying memory},\ }\href@noop {} {\bibfield  {journal} {\bibinfo  {journal}
  {arXiv preprint arXiv:2509.01806}\ } (\bibinfo {year} {2025})}\BibitemShut
  {NoStop}%
\bibitem [{\citenamefont {Kenkre}(2007)}]{kenkre2007analytic}%
  \BibitemOpen
  \bibfield  {author} {\bibinfo {author} {\bibfnamefont {V.}~\bibnamefont
  {Kenkre}},\ }\bibfield  {title} {\bibinfo {title} {Analytic formulation,
  exact solutions, and generalizations of the elephant and the {A}lzheimer
  random walks},\ }\href@noop {} {\bibfield  {journal} {\bibinfo  {journal}
  {arXiv preprint arXiv:0708.0034}\ } (\bibinfo {year} {2007})}\BibitemShut
  {NoStop}%
\bibitem [{\citenamefont {Baur}\ and\ \citenamefont
  {Bertoin}(2016)}]{baur2016elephant}%
  \BibitemOpen
  \bibfield  {author} {\bibinfo {author} {\bibfnamefont {E.}~\bibnamefont
  {Baur}}\ and\ \bibinfo {author} {\bibfnamefont {J.}~\bibnamefont {Bertoin}},\
  }\bibfield  {title} {\bibinfo {title} {Elephant random walks and their
  connection to {P}{\'o}lya-type urns},\ }\href@noop {} {\bibfield  {journal}
  {\bibinfo  {journal} {Physical Review E}\ }\textbf {\bibinfo {volume} {94}},\
  \bibinfo {pages} {052134} (\bibinfo {year} {2016})}\BibitemShut {NoStop}%
\bibitem [{\citenamefont {Coletti}\ \emph {et~al.}(2017)\citenamefont
  {Coletti}, \citenamefont {Gava},\ and\ \citenamefont
  {Sch{\"u}tz}}]{coletti2017central}%
  \BibitemOpen
  \bibfield  {author} {\bibinfo {author} {\bibfnamefont {C.~F.}\ \bibnamefont
  {Coletti}}, \bibinfo {author} {\bibfnamefont {R.}~\bibnamefont {Gava}},\ and\
  \bibinfo {author} {\bibfnamefont {G.~M.}\ \bibnamefont {Sch{\"u}tz}},\
  }\bibfield  {title} {\bibinfo {title} {Central limit theorem and related
  results for the elephant random walk},\ }\href@noop {} {\bibfield  {journal}
  {\bibinfo  {journal} {Journal of Mathematical Physics}\ }\textbf {\bibinfo
  {volume} {58}} (\bibinfo {year} {2017})}\BibitemShut {NoStop}%
\bibitem [{\citenamefont {Laulin}(2022)}]{laulin2022introducing}%
  \BibitemOpen
  \bibfield  {author} {\bibinfo {author} {\bibfnamefont {L.}~\bibnamefont
  {Laulin}},\ }\bibfield  {title} {\bibinfo {title} {Introducing smooth amnesia
  to the memory of the elephant random walk},\ }\href@noop {} {\bibfield
  {journal} {\bibinfo  {journal} {Electronic Communications in Probability}\
  }\textbf {\bibinfo {volume} {27}},\ \bibinfo {pages} {1} (\bibinfo {year}
  {2022})}\BibitemShut {NoStop}%
\bibitem [{\citenamefont {Gut}\ and\ \citenamefont
  {Stadtm{\"u}ller}(2023)}]{gut2023elephant}%
  \BibitemOpen
  \bibfield  {author} {\bibinfo {author} {\bibfnamefont {A.}~\bibnamefont
  {Gut}}\ and\ \bibinfo {author} {\bibfnamefont {U.}~\bibnamefont
  {Stadtm{\"u}ller}},\ }\bibfield  {title} {\bibinfo {title} {Elephant random
  walks; a review},\ }\href@noop {} {\bibfield  {journal} {\bibinfo  {journal}
  {Ann. Univ. Sci. Budapest. Sect. Comput}\ }\textbf {\bibinfo {volume} {54}},\
  \bibinfo {pages} {171} (\bibinfo {year} {2023})}\BibitemShut {NoStop}%
\bibitem [{\citenamefont {Linkenkaer-Hansen}\ \emph {et~al.}(2001)\citenamefont
  {Linkenkaer-Hansen}, \citenamefont {Nikouline}, \citenamefont {Palva},\ and\
  \citenamefont {Ilmoniemi}}]{linkenkaer2001long}%
  \BibitemOpen
  \bibfield  {author} {\bibinfo {author} {\bibfnamefont {K.}~\bibnamefont
  {Linkenkaer-Hansen}}, \bibinfo {author} {\bibfnamefont {V.~V.}\ \bibnamefont
  {Nikouline}}, \bibinfo {author} {\bibfnamefont {J.~M.}\ \bibnamefont
  {Palva}},\ and\ \bibinfo {author} {\bibfnamefont {R.~J.}\ \bibnamefont
  {Ilmoniemi}},\ }\bibfield  {title} {\bibinfo {title} {Long-range temporal
  correlations and scaling behavior in human brain oscillations},\ }\href@noop
  {} {\bibfield  {journal} {\bibinfo  {journal} {Journal of Neuroscience}\
  }\textbf {\bibinfo {volume} {21}},\ \bibinfo {pages} {1370} (\bibinfo {year}
  {2001})}\BibitemShut {NoStop}%
\bibitem [{\citenamefont {Leighton}\ and\ \citenamefont
  {Lynn}(2025)}]{leighton2025main}%
  \BibitemOpen
  \bibfield  {author} {\bibinfo {author} {\bibfnamefont {M.~P.}\ \bibnamefont
  {Leighton}}\ and\ \bibinfo {author} {\bibfnamefont {C.~W.}\ \bibnamefont
  {Lynn}},\ }\bibfield  {title} {\bibinfo {title} {Decomposing non-markovian
  history dependence},\ }\href@noop {} {\bibfield  {journal} {\bibinfo
  {journal} {arXiv preprint arXiv:2512.13933}\ } (\bibinfo {year}
  {2025})}\BibitemShut {NoStop}%
\bibitem [{\citenamefont {Conway}(2012)}]{conway2012functions}%
  \BibitemOpen
  \bibfield  {author} {\bibinfo {author} {\bibfnamefont {J.~B.}\ \bibnamefont
  {Conway}},\ }\href@noop {} {\emph {\bibinfo {title} {Functions of one complex
  variable II}}},\ Vol.\ \bibinfo {volume} {159}\ (\bibinfo  {publisher}
  {Springer Science \& Business Media},\ \bibinfo {year} {2012})\BibitemShut
  {NoStop}%
\bibitem [{\citenamefont {Cover}\ and\ \citenamefont
  {Thomas}(2006)}]{Cover2006_Elements}%
  \BibitemOpen
  \bibfield  {author} {\bibinfo {author} {\bibfnamefont {T.~M.}\ \bibnamefont
  {Cover}}\ and\ \bibinfo {author} {\bibfnamefont {J.~A.}\ \bibnamefont
  {Thomas}},\ }\href@noop {} {\emph {\bibinfo {title} {Elements of Information
  Theory}}},\ \bibinfo {edition} {2nd}\ ed.\ (\bibinfo  {publisher}
  {Wiley-Interscience},\ \bibinfo {address} {Hoboken, NJ},\ \bibinfo {year}
  {2006})\BibitemShut {NoStop}%
\bibitem [{\citenamefont {Flajolet}\ and\ \citenamefont
  {Odlyzko}(1990)}]{flajolet1990singularity}%
  \BibitemOpen
  \bibfield  {author} {\bibinfo {author} {\bibfnamefont {P.}~\bibnamefont
  {Flajolet}}\ and\ \bibinfo {author} {\bibfnamefont {A.}~\bibnamefont
  {Odlyzko}},\ }\bibfield  {title} {\bibinfo {title} {Singularity analysis of
  generating functions},\ }\href@noop {} {\bibfield  {journal} {\bibinfo
  {journal} {SIAM Journal on Discrete Mathematics}\ }\textbf {\bibinfo {volume}
  {3}},\ \bibinfo {pages} {216} (\bibinfo {year} {1990})}\BibitemShut {NoStop}%
\bibitem [{\citenamefont {Flajolet}(2003)}]{flajolet2003singular}%
  \BibitemOpen
  \bibfield  {author} {\bibinfo {author} {\bibfnamefont {P.}~\bibnamefont
  {Flajolet}},\ }\bibfield  {title} {\bibinfo {title} {Singular
  combinatorics},\ }\href@noop {} {\bibfield  {journal} {\bibinfo  {journal}
  {arXiv preprint math/0304465}\ } (\bibinfo {year} {2003})}\BibitemShut
  {NoStop}%
\bibitem [{\citenamefont {Vaswani}\ \emph {et~al.}(2017)\citenamefont
  {Vaswani}, \citenamefont {Shazeer}, \citenamefont {Parmar}, \citenamefont
  {Uszkoreit}, \citenamefont {Jones}, \citenamefont {Gomez}, \citenamefont
  {Kaiser},\ and\ \citenamefont {Polosukhin}}]{vaswani2017attention}%
  \BibitemOpen
  \bibfield  {author} {\bibinfo {author} {\bibfnamefont {A.}~\bibnamefont
  {Vaswani}}, \bibinfo {author} {\bibfnamefont {N.}~\bibnamefont {Shazeer}},
  \bibinfo {author} {\bibfnamefont {N.}~\bibnamefont {Parmar}}, \bibinfo
  {author} {\bibfnamefont {J.}~\bibnamefont {Uszkoreit}}, \bibinfo {author}
  {\bibfnamefont {L.}~\bibnamefont {Jones}}, \bibinfo {author} {\bibfnamefont
  {A.~N.}\ \bibnamefont {Gomez}}, \bibinfo {author} {\bibfnamefont
  {{\L}.}~\bibnamefont {Kaiser}},\ and\ \bibinfo {author} {\bibfnamefont
  {I.}~\bibnamefont {Polosukhin}},\ }\bibfield  {title} {\bibinfo {title}
  {Attention is all you need},\ }\href@noop {} {\bibfield  {journal} {\bibinfo
  {journal} {Advances in Neural Information Processing Systems}\ }\textbf
  {\bibinfo {volume} {30}} (\bibinfo {year} {2017})}\BibitemShut {NoStop}%
\end{thebibliography}%

\section*{End Matter}

\section*{Appendix A: Exact calculations for exponential dependence}

Here we detail exact calculations for the copy model with exponential history dependence.

\subsection*{A.1 \,\,\,Autocorrelations}
We compute the autocorrelation function $C(k)$ exactly using $C(0)=1$ and the recurrence relation
\begin{equation}
C(k) = (2\alpha-1)\sum_{k'=1}^\infty C(k-k')\rho(k'),
\end{equation}
which is valid for $k>0$.
We use an ansatz $C(k) = AB^{k-1}$ for $k>0$, and can thus write the recurrence relation for $k>0$ as
\begin{equation}
\begin{aligned}\label{eq:expCkeqn1}
AB^{-k} & = (2\alpha-1)AqB^{k-1}\sum_{k'=1}^{k-1}B^{-k'}(1-q)^{k'-1}\\
& + (2\alpha-1)q(1-q)^{k-1}\\
& + (2\alpha-1)AqB^{-k-1}\sum_{k'=k+1}^\infty B^{k'}(1-q)^{k'-1}.
\end{aligned}
\end{equation}
Here we used $C(0)=0$ and $C(-k)=C(k)$, splitting the infinite sum in the recurrence into contributions for $k'<k$, $k'=k$, and $k'>k$.
Computing the two sums, and rearranging, we can rewrite Eq.~\eqref{eq:expCkeqn1} as
\begin{equation}
\begin{aligned}
& AB^{k-1} \left[\frac{1}{q(2\alpha-1)} + \frac{1}{1-q-B}\right] \\
& = (1-q)^{k-1}\left[1 + \frac{A}{1-q-B} + \frac{A(1-q)}{1-B(1-q)}\right].
\end{aligned}
\end{equation}
Here we have grouped terms involving exponentials with bases $B$ and $(1-q)$ on different sides. Since this equality must hold for all $k>0$, it follows that both sides must be exactly zero. We thus obtain two equations that can be solved simultaneously for $A$ and $B$, which yields
\begin{subequations}
\begin{align}
A & = \underbrace{(2\alpha-1) q \left[1 + \frac{(2\alpha-1)(1-q)}{4 + 4\alpha q - 4\alpha - 3q}\right]}_{\equiv\gamma},\\
B & = 1-2q+2\alpha q,
\end{align}
\end{subequations}
and thus recovering Eq.~\eqref{eq:expCk}.

\subsection*{A.2 \,\,\,Collective variable}
We compute the variance of the collective variable $Y(t)$ beginning with the Markovian update rule
\begin{equation}
Y(t+1) = \begin{cases} (1-q) Y(t) + q & \mathrm{with}\,p(+)=\,\frac{1}{2} + \frac{2\alpha-1}{2}Y(t),\\
(1-q) Y(t) - q & \mathrm{with}\,p(-)=\,\frac{1}{2} - \frac{2\alpha-1}{2}Y(t). \end{cases}
\end{equation}
Since the process is unbiased, $\langle Y(t)\rangle=0$, and thus the variance is given by $\langle Y(t)^2\rangle$. We compute the variance as
\begin{subequations}
\begin{align}
& \left\langle Y(t+1)^2\right\rangle_{p(Y(t+1))} \nonumber \\
& = \left\langle \left\langle Y(t+1)^2\right\rangle_{p(Y(t+1)|Y(t))}\right\rangle_{p(Y(y))}\\
& = \left\langle \left[(1-q)Y(t)+q\right]^2\left[\frac{1}{2}+\frac{2\alpha-1}{2}Y(t)\right]\right.\nonumber\\
& + \left\langle \left[(1-q)Y(t)-q\right]^2\left[\frac{1}{2}-\frac{2\alpha-1}{2}Y(t)\right]\right\rangle_{p(Y(t))}\\
& = q^2 + \left[(1-q)^2 + 2q(1-q)(2\alpha-1)\right]\langle Y(t)^2\rangle_{p(Y(t))}.
\end{align}
\end{subequations}
Since the process reaches a steady state, the statistics of $Y(t+1)$ and $Y(t)$ are identical; in particular $\langle Y(t+1)^2\rangle = \langle Y(t)^2\rangle$. We can thus solve for the variance, obtaining
\begin{equation}
\left\langle Y(t)^2\right\rangle = \frac{q}{4 + 4\alpha q - 4\alpha - 3q}.
\end{equation}

\subsection*{A.3 \,\,\,Markovian (first-order) dynamical information}

Using the variance of $Y$, we can then exactly compute the first-order dynamical information $I_1$. Since $h_0=\ln2$ (because the process is unbiased), we need only compute $h_1$. To compute $h_1$, we need the conditional distribution $p(x_t|x_{t-1}$), which, using the collective variable, we can write as
\begin{subequations}
\begin{align}
& p(x_t|x_{t-1})\nonumber\\
& = \frac{p(x_{t},x_{t-1})}{p(x_{t-1})}\\
& = \frac{ \left\langle p(x_t|x_{t-1},Y(t-1))p(x_{t-1}|Y(t-1))\right\rangle_{p(Y(t-1))}}{p(x_{t-1})}.\label{eq:marginalconditional}
\end{align}
\end{subequations}
Here we have used the collective variable to simplify the conditional distributions describing history dependence. Using Eq.~\eqref{eq:condprob}, these can be written as
\begin{equation}\begin{aligned}
& p(x_t|x_{t-1},Y(t-1)) \\
& = \frac{1}{2} + \frac{1}{2}(2\alpha-1)x_t\left[(1-q)Y(t-1) + qx_{t-1}\right],
\end{aligned}
\end{equation}
and
\begin{equation}
p(x_{t-1}|Y(t-1)) = \frac{1}{2} + \frac{1}{2}(2\alpha-1)Y(t-1)x_{t-1}.
\end{equation}
Multiplying these two conditional distributions, evaluating the average over $p(Y(t-1))$ and identifying the variance $\langle Y^2\rangle$, and dividing by $p(x_{t-1})=1/2$ allows us to evaluate Eq.~\eqref{eq:marginalconditional} as
\begin{equation}
\begin{aligned}
& p(x_t|x_{t-1}) \\
& = \frac{1}{2} + \frac{1}{2}x_tx_{t-1}\underbrace{(2\alpha-1) q \left[1 + \frac{(2\alpha-1)(1-q)}{4 + 4\alpha q - 4\alpha - 3q}\right]}_{\equiv\gamma}.
\end{aligned}
\end{equation}
Here we have defined $\gamma$ to simplify the notation. The conditional entropy $h_1$ can then be evaluated by summing over all combinations of $x_t$ and $x_{t-1}$,
\begin{subequations}
\begin{align}
h_1 & \equiv H[x_t|x_{t-1}]\\
& = -\sum_{x_t=\pm1}\sum_{x_{t-1}=\pm1} p(x_{t-1})p(x_t|x_{t-1})\ln p(x_t|x_{t-1})\\
& = -\left(\frac{1}{2} + \frac{\gamma}{2}\right)\ln\left(\frac{1}{2} + \frac{\gamma}{2}\right) - \left(\frac{1}{2} - \frac{\gamma}{2}\right)\ln\left(\frac{1}{2} - \frac{\gamma}{2}\right).
\end{align}
\end{subequations}
Finally, we can evaluate the first-order dynamical information as
\begin{subequations}
\begin{align}
I_1 & \equiv h_0 - h_1\\
& = \ln2 \\
& + \left(\frac{1}{2} + \frac{\gamma}{2}\right)\ln\left(\frac{1}{2} + \frac{\gamma}{2}\right) + \left(\frac{1}{2} - \frac{\gamma}{2}\right)\ln\left(\frac{1}{2} - \frac{\gamma}{2}\right)\nonumber\\
& = \Phi(\gamma).
\end{align}
\end{subequations}
This gives our result, Eq.~\eqref{eq:exp_copy_I1}.

\section*{Appendix B: Autocorrelation scaling for power-law dependence}

\subsection*{B.1\,\,\, Long-time limits from a transfer theorem}
Here we derive the long-term scaling of the autocorrelation for power-law history dependence. We begin with the autocorrelation generating function, which is
\begin{equation}
\tilde{C}(z) = \frac{1 + (2\alpha-1)A(z)}{1 - (2\alpha-1)\mathrm{Li}_s(z)/\zeta(s)}.
\end{equation}
For noninteger $s>1$ and $u=1-z$, the autocorrelation generating function can be expanded around $u=0$ as
\begin{equation}
\begin{aligned}
\tilde{C}(z) \sim & \mathcal{O}(1) + \left[ \mathcal{O}(u) + \mathcal{O}(u^2) + \hdots \right] \\
& + \left[ \mathcal{O}(u^{s-1}) + \mathcal{O}(u^{s}) + \mathcal{O}(u^{s+1}) + \hdots \right].
\end{aligned}
\end{equation}
For integer $s>1$, we instead get
\begin{equation}
\begin{aligned}
\tilde{C}(z) \sim & \mathcal{O}(1) + \left[ \mathcal{O}(u) + \mathcal{O}(u^2) + \hdots \right]\\
& + \left[ \mathcal{O}(u^{s-1}) + \mathcal{O}(u^{s}) + \hdots \right]\left[\mathcal{O}(1) + \mathcal{O}\left(\ln u\right)\right].
\end{aligned}
\end{equation}
Thus for all $s>2$ the dominant term is $\mathcal{O}(u)$, while for $s=2$ the dominant term is instead $\mathcal{O}\left(u\ln u\right)$, and for $s<2$ the term $\mathcal{O}(u^{s-1})$ dominates. Note that the latter two are non-analytic singularities, while the former is analytic. For $s>2$ the dominant non-analytic term is either $\mathcal{O}\left(u^{s-1}\right)$ for noninteger $s$, or $\mathcal{O}\left(u^{s-1}\ln u\right)$ for integer $s$.

To determine the long-time scaling of the autocorrelations, we use the Tauberian transfer theorem proven by Flajolet and Odlyzko~\cite{flajolet1990singularity,flajolet2003singular}, which states that if the generating function $g(z)$ of a discrete function $f(k)$ has non-analytically singular scaling for $u=1-z$, $u\to 0$ as
\begin{equation}
g(u)\sim u^{-\beta} \left(\ln u\right)^\eta,
\end{equation}
then as $k\to\infty$ we have
\begin{equation}
f(k)\sim k^{\beta-1}\left(\ln k\right)^\eta.
\end{equation}
Crucially here we must use the dominant non-analytic term. 

Thus for general $s$, the autocorrelations in the long-time limit must scale as
\begin{equation}
C(k)\sim\begin{cases} k^{-s}\ln k, & \mathrm{integer}\,\,s,\\
k^{-s}, & \mathrm{noninteger}\,\,s.
\end{cases}
\end{equation}
The finite-time exponential scaling for $s>2$ arises due to the $\mathcal{O}(u)$ term in the expansion, which produces a simple pole near $z=1$.

\subsection*{B.2\,\,\, Exponential to power law transition}

For $s>2$, we showed in the main text that for small $k$ the autocorrelations decay exponentially as $C(k)\propto z_c^{-k}$, while in the previous Appendix subsection we showed that for large $k$ they scale as $C(k)\propto k^{-s}$ (here we ignore the logarithmic correction at integer $k$, which does not significantly effect these results). To determine the point of transition, we set
\begin{subequations}
\begin{align}
C_\mathrm{exp}(k) & = A_\mathrm{exp}\, z_c^{-k},\\
C_\mathrm{pl}(k) & = A_\mathrm{pl}\, k^{-s}.
\end{align}
\end{subequations}
We expect both proportionality constants $A_\mathrm{exp}$ and $A_\mathrm{pl}$ to be $\mathcal{O}(1)$, so set them to be equal initially. Solveing for $C_\mathrm{exp}(k) = C_\mathrm{pl}(k)$, we obtain
\begin{equation}
k^*= -\frac{s}{\ln z_c}W_{-1}\left(-\frac{\ln z_c}{s}\right).
\end{equation}
Here $W_{-1}(z)$ is the $-1$ branch of the Lambert W function. Both the $0$ and $-1$ branches give valid real solutions for $k^*$, but the $W_0(z)$ branch gives $k^*\approx 1$ and thus does not correspond to the crossover we seek. The $W_{-1}(z)$ branch, by contrast, gives results consistent with numerical simulations.

More generally we will have $A_\mathrm{exp}\neq A_\mathrm{pl}$. To determine how much this affects our results, we expand $k^*$ around $A_\mathrm{exp} = A_\mathrm{pl}$; to first order we find
\begin{equation}
\begin{aligned}
& k^* = \\
& -\frac{s}{\ln z_c}W_{-1}\left(-\frac{\ln z_c}{s}\right)\left[1 - \frac{\frac{A_\mathrm{exp}- A_\mathrm{pl}}{A_\mathrm{pl}}}{s\left(1 + W_{-1}\left(-\frac{\ln z_c}{s}\right)\right)} \right] \\
& + \mathcal{O}\left(\frac{A_\mathrm{exp}- A_\mathrm{pl}}{A_\mathrm{pl}}\right)^2.
\end{aligned}
\end{equation}
For $\alpha=0.9$ and $s=4$, for example, we then have
\begin{equation}
k^* \approx 88.3\left[ 1 + 0.07*\left(\frac{A_\mathrm{exp}- A_\mathrm{pl}}{A_\mathrm{pl}}\right)\right],
\end{equation}
which shows that the correction will be small for even moderate differences between $A_\mathrm{exp}$ and $A_\mathrm{pl}$. More generally, numerical exploration shows that even for order-of-magnitude differences between $A_\mathrm{exp}$ and $A_\mathrm{pl}$ we still shouldn't see $k^*$ smaller than $\mathcal{O}(100)$.

\section*{Appendix C: Exact conditional entropy for the ERW}

\subsection*{C.1 \,\,\,First-order dynamical information}
As in the exponential case, we can use the known variance of $Y$ to exactly compute the first-order dynamical information $I_1$ in the long-time limit. Since $h_0=\ln2$ (because the process is unbiased), we need only compute $h_1$, which we can compute from the conditional distribution $p(x_t|x_{t-1}$), which, following our arguments in Appendix A, can be written as 
\begin{subequations}
\begin{align}
& p(x_t|x_{t-1})\nonumber\\
& = \frac{ \left\langle p(x_t|x_{t-1},Y(t-1))p(x_{t-1}|Y(t-1))\right\rangle_{p(Y(t-1))}}{p(x_{t-1})}.
\end{align}
\end{subequations}
Using Eq.~\eqref{eq:condprob}, the two conditional distributions inside the average can be written as
\begin{equation}\begin{aligned}
& p(x_t|x_{t-1},Y(t-1)) \\
& = \frac{1}{2} + \frac{1}{2}(2\alpha-1)x_t\left[\frac{t-1}{t}Y(t-1) + \frac{1}{t}x_{t-1}\right],
\end{aligned}
\end{equation}
and
\begin{equation}
p(x_{t-1}|Y(t-1)) = \frac{1}{2} + \frac{1}{2}(2\alpha-1)x_{t-1} Y(t-1).
\end{equation}
Multiplying these two conditional distributions, evaluating the average over $p(Y(t-1))$ and identifying the variance $\langle Y(t-1)^2\rangle$, and dividing by $p(x_{t-1})=1/2$ allows us to evaluate the final conditional probability as
\begin{equation}
\begin{aligned}
& p(x_t|x_{t-1}) \\
& = \frac{1}{2} + \frac{1}{2}x_tx_{t-1}\underbrace{(2\alpha-1) \frac{1}{t} \left[1 + (2\alpha-1)(t-1)\left\langle Y(t-1)^2\right\rangle\right]}_{\equiv\delta}.
\end{aligned}
\end{equation}
Here we have defined $\delta$ to simplify the notation. As before, the conditional entropy $h_1$ can then be evaluated by summing over all combinations of $x_t$ and $x_{t-1}$. Computing the first-order dynamical information, we obtain
\begin{equation}
I_1  = \Phi(\delta).
\end{equation}
This gives our result, Eq.~\eqref{eq:ERW_I1}.

\vspace{0.5cm}
\subsection*{C.2 \,\,\,Recursive formula for higher orders}
The approach used above suggests a recursive approach to approximately compute higher-order dynamical information, which simply requires computing the conditional entropies $h_k$ for arbitrary $k$. Each $h_k$ requires computing the marginal distribution over the trajectory including states from $t-k$ to $t$, as well as the conditional distribution of $x_t$ given states from $x_{t-1}$ to $x_{t-k}$. For a given $k$ these can be written recursively in terms of the marginal and conditional distributions for smaller $k$. In particular we need three key equations, which are:
\begin{widetext}
\begin{subequations}
\begin{align}
p(x_t,x_{t-1},\hdots ,x_{t-k}) & = \left\langle  \prod_{i=0}^{k-1}p(x_{t-i}|x_{t-i-1},\hdots ,x_{t-k},Y(t-k)) p(x_{t-k}|Y(t-k)) \right\rangle_{p(Y(t-k))},\\
p(x_{t-i}|x_{t-i-1},\hdots ,x_{t-k},Y(t-k)) & = \frac{1}{2} + \frac{1}{2}(2\alpha-1)x_{t-i}\left[ \frac{t-k}{t-i}Y(t-k) + \frac{1}{t-i}\sum_{j=1}^{k-i}x_j \right],\\
p(x_t|x_{t-1},\hdots ,x_{t-k})& = \frac{p(x_t,x_{t-1},\hdots ,x_{t-k})}{p(x_{t-k})\prod_{i=0}^{k-1}p(x_{t-i}|x_{t-i-1},\hdots ,x_{t-k})}.
\end{align}
\end{subequations}
\end{widetext}
The first computes the marginal probability using a product conditional on the full history, which can be simplified using the collective variable $Y(t-k)$, then marginalized by averaging over $p(Y(t-k))$. The second equation gives the explicit history dependence written in terms of the collective variable. The third equation then writes the key conditional distribution dependent on history only up to $x_{t-k}$ in terms of only the marginal distribution and conditionals over less history. Thus given knowledge of $p(x_{t-k+1},x_{t-k})$, which we computed in the previous section in the process of deriving the first-order dynamical information, these equations can be used to recursively compute higher-order marginal and conditional probabilities. 

The average over $p(Y(t-k))$ can be evaluated approximately using the known Gaussian nature of the collective variable for $\alpha\leq3/4$, by approximating the sum as an integral over the Gaussian distribution, which will converge to the exact result in the long-time limit. Since the distribution of the collective variable is not known for $\alpha>3/4$, this approach does not work in that regime.

Once the marginal and conditional probabilities have been computed, the conditional entropy can then be computed as
\begin{equation}\label{eq:h_k_recursion}
h_k = -\sum_{\{x_j\}_{t-k}^t}p(x_t,x_{t-1},\hdots ,x_{t-k})\ln p(x_t|x_{t-1},\hdots ,x_{t-k}).
\end{equation}
By computing the entropies $h_k$ recursively to larger and larger $k$, it is then straightforward to compute the dynamical information by taking differences of consecutive conditional entropies. Since for each conditional entropy we must enumerate all possible trajectories of length $k+1$, the computational cost of this recursive algorithm scales exponentially with $k$. In practice this approach allows for approximate computation of the dynamical information up to $k=20$ in less than a day on a typical laptop.

\end{document}